\documentclass[useAMS,usenatbib,referee]{mnras}
\usepackage{graphicx}
\title[GALAH Observational Overview]{The GALAH Survey: Observational Overview and {\it Gaia} DR1 companion}
\author[S. L. Martell et al.]{S. L. Martell$^{1}$\thanks{email: s.martell@unsw.edu.au}, S. Sharma$^{2}$, S. Buder$^{3}$, L. Duong$^{4}$, K. J. Schlesinger$^{4}$, \newauthor J. Simpson$^{5}$, K. Lind$^{3,6}$, M. Ness$^{3}$, J. P. Marshall$^{1,7}$, M. Asplund$^{4}$, \newauthor J. Bland-Hawthorn$^{2}$, A. R. Casey$^{8}$, G. De Silva$^{2,5}$, K. C. Freeman$^{4}$,  J. Kos$^{2}$, \newauthor J. Lin$^{4}$, D. B. Zucker$^{5,9,10}$, T. Zwitter$^{11}$, B. Anguiano$^{9,10}$, C. Bacigalupo$^{9,10}$, \newauthor D. Carollo$^{12}$, L. Casagrande$^{4}$, G. S. Da Costa$^{4}$, J. Horner$^{7,13}$, D. Huber$^{2}$, \newauthor E. A. Hyde$^{5,14}$, P. R. Kafle$^{15}$,  G. F. Lewis$^{2}$, D. Nataf$^{4}$, D. Stello$^{2}$, \newauthor C. G. Tinney$^{1,7}$, F. G. Watson$^{5}$, R. Wittenmyer$^{13,1,7}$\\
$^1$School of Physics, University of New South Wales, Sydney NSW 2052, Australia\\
$^2$Sydney Institute for Astronomy, School of Physics, A28, The University of Sydney, Sydney NSW 2006,  Australia\\
$^3$Max-Planck-Institut f{\"u}r Astronomie, K{\"o}nigstuhl 17, 69117 Heidelberg, Germany\\
$^4$Research School of Astronomy \& Astrophysics, Australian National University, Canberra ACT 2611, Australia\\
$^5$Australian Astronomical Observatory, North Ryde NSW 2113, Australia\\
$^6$Department of Physics and Astronomy, Uppsala University, Box 516, SE-751 20 Uppsala, Sweden\\
$^7$Australian Centre for Astrobiology, University of New South Wales, Sydney NSW 2052, Australia\\
$^8$Institute of Astronomy, University of Cambridge, Cambridge, CB3 0HA, UK\\
$^9$Department of Physics and Astronomy, Macquarie University, Sydney NSW 2109, Australia\\
$^{10}$Research Centre in Astronomy, Astrophysics and Astrophotonics, Macquarie University, Sydney NSW 2109, Australia\\
$^{11}$Faculty of Mathematics and Physics, University of Ljubljana, Jadranska 19, 1000 Ljubljana, Slovenia\\
$^{12}$Department of Physics and JINA Center for the Evolution of the Elements, University of Notre Dame, Notre Dame, IN 46556, USA\\
$^{13}$Computational Engineering and Science Research Centre, University of Southern Queensland, Towoomba QLD 4350, Australia\\
$^{14}$Western Sydney University, Locked Bag 1797, Penrith South DC, NSW 1797, Australia\\
$^{15}$International Centre for Radio Astronomy Research,The University of Western Australia, WA 6009, Australia}

\begin{document}

\date{Accepted; Received}
\pagerange{\pageref{firstpage}--\pageref{lastpage}} \pubyear{2016}

\maketitle

\begin{abstract}
The Galactic Archaeology with HERMES (GALAH) Survey is a massive observational project to trace the Milky Way's history of star formation, chemical enrichment, stellar migration and minor mergers. Using high-resolution (R$\simeq$28,000) spectra taken with the High Efficiency and Resolution Multi-Element Spectrograph (HERMES) instrument at the Anglo-Australian Telescope (AAT), GALAH will determine stellar parameters and abundances of up to $29$ elements for up to one million stars. Selecting targets from a colour-unbiased catalogue built from 2MASS, APASS and UCAC4 data,  we expect to observe dwarfs at 0.3 to 3~kpc and giants at 1 to 10~kpc. This enables a thorough local chemical inventory of the Galactic thin and thick disks, and also captures smaller samples of the bulge and halo. In this paper we present the plan, process and progress as of early 2016 for GALAH survey observations. In our first two years of survey observing we have accumulated the largest high-quality spectroscopic data set at this resolution, over 200,000 stars. We also present the first public GALAH data catalogue: stellar parameters (T$_{\rm eff}$, log(g), [Fe/H], [$\alpha$/Fe]), radial velocity, distance modulus and reddening for 10680 observations of 9860 {\it Tycho-2} stars that may be included in the first {\it Gaia} data release. \end{abstract}

\begin{keywords}
stars: abundances -- Galaxy: disc -- Galaxy: formation -- Galaxy: evolution -- Galaxy: stellar content
\end{keywords}

\section{Introduction}
Massive observational surveys are an increasingly important force in astronomy. In particular, spectroscopic stellar surveys are revolutionising our understanding of Galactic structure and evolution (e.g., \citealt{H08}; \citealt{RB13}; \citealt{HHB14}; \citealt{HBH15}; \citealt{MFR16}). As in many areas of astronomical research, this development is driven by technology. Efficient methods for accurately positioning many optical fibres at telescope focal planes are enabling an increasing number of observatories to add highly multiplexed high-resolution spectrographs to their instrument suites (e.g., \citealt{CZC12}; \citealt{STK15}).

The Galactic Archaeology with HERMES (GALAH) Survey\footnote{http://galah-survey.org} is a high-resolution spectroscopic survey that is exploring the chemical and dynamical history of the Milky Way, with particular focus on the disk. GALAH aims to collect a comprehensive data set, in terms of both sample size and detail, with abundances for as many as 29 elements (Li, C, O, Na, Mg, Al, Si, K, Ca, Sc, Ti, V, Cr, Mn, Fe, Co, Ni, Cu, Zn, Rb, Sr, Y, Zr, Ru, Ba, La, Ce, Nd, Eu) for each target. Our overall science goal is to carry out chemical tagging (e.g., \citealt{FBH02}; \citealt{DSS06}; \citealt{BHK10}) within this data set, identifying stars that formed at the same time and place by matching their abundance patterns. A thorough explanation of the GALAH survey science goals is given in \citet{DSF15}.

The project of chemical tagging in the Galactic disk demands a very large data set. Given the observational selection for GALAH targets (discussed in Section 3 below), we anticipate that roughly $75\%$ of stars observed by GALAH will belong to the thin disk and $24\%$ to the thick disk, with smaller numbers of nearby halo stars and bright red giants in the bulge making up the rest of the sample. From theoretical explorations of clustered star formation (e.g., \citealt{BHKF10}; \citealt{FK14}), we expect there to be stars from a large number of unique star-forming events (``initial star-forming groups") mixed throughout both the thin and thick disks. The number of these groups in the disk, and the number of stars per group, will depend on the initial mass function and maximum mass of each group. \citet{TCG15} predicted that a survey of 10$^5$ stars can expect to observe 10 stars per group down to an initial mass limit of $\sim 10^6 M_{\odot}$, while a survey of 10$^6$ stars would capture 10 stars per group down to a group mass of $\sim 10^5 M_{\odot}$. \citet{TCR16} took data for 13,000 stars from the Apache Point Observatory Galaxy Evolution Experiment (APOGEE; \citealt{MSF15}; \citealt{HSJ15}) Survey from the twelfth data release of the Sloan Digital Sky Survey (DR12, \citealt{Alam15}). By analysing the ``clumpiness" of the chemical abundance data rather than carrying out strict chemical tagging, they were able to rule out the presence of an initial star-forming group in the thick disk with a mass greater than 10$^7 M_{\odot}$.

Our observational program must therefore collect enough stars from each initial star-forming group, and derive precise enough stellar parameters and elemental abundances, to confidently apply chemical tags to them. Since the observing time for GALAH is allocated through the competitive time allocation process of the 3.9m Anglo-Australian Telescope (AAT), our observing strategy must provide this large, high-quality sample in a reasonable amount of observing time. This paper describes the balance between sample size, signal-to-noise ratio (SNR) and observing time that has been designed into our observational program. Section 2 outlines the capabilities of the HERMES spectrograph and Two Degree Field (2dF) fibre positioner, Section 3 discusses our target selection for the main survey, Section 4 describes the observing procedure, Section 5 discusses the Pilot Survey, Section 6 describes the K2-HERMES program, Section 7 presents observing progress through January 2016 (the end of AAT observing semester 15B), Section 8 discusses our potential synergies with other large Galactic survey programs, and Section 9 presents the GALAH-TGAS catalogue. The overlap between GALAH and {\it Gaia} is an extremely important data set. GALAH stars are all in the magnitude range (12$\le V \le$ 14) for which {\it Gaia} parallaxes and proper motions will be at their best and most complete. Ultimately GALAH will be able to contribute elemental abundances for a large number of stars with high-precision {\it Gaia} data, forming a very powerful resource for studying Galactic chemodynamics.

\section{Instrumentation: the HERMES Spectrograph and the 2dF Fibre Positioner}
The GALAH survey collects all of its data with the HERMES spectrograph at the AAT. While HERMES is an AAT facility instrument, it was specifically designed to undertake a large Galactic archaeology survey \citep{F12}. The instrumental requirements for efficiency, wavelength range and spectral resolution were therefore focused on producing spectra rich in information about stellar parameters and chemical abundances, across a wide range of stellar effective temperature, surface gravity and overall metallicity. Details of HERMES design and integration can be found in \citet{BJB10}, \citet{BCG11}, \citet{HAB12} and \citet{FBH14}, and the as-built performance is discussed in \citet{SAA15}.

HERMES has four non-contiguous optical bandpasses covering a total of $\sim$1000$\hbox{\AA}$, with wavelength ranges chosen to maximise the information captured for determining stellar parameters and abundances (see Table 1). A series of dichroic beam splitters divides incoming light into the four channels, with a separate volume phase holographic grating and camera for each. The cameras have independent shutters, and can be given different exposure times. This feature is mainly used during flat-field exposures, when the exposure times are set to ($180$, $180$, $120$, $90$) seconds in the blue, green, red and infrared channels, respectively, to deliver relatively even count levels (averaging between $5000$ and $17000$ counts across all fibres and wavelengths in the raw data) in all four cameras. Spectral resolution, as measured from ThXe arc lamp exposures, is $R \sim28,000$. A more detailed analysis of the spectrograph resolving power as a function of position on the detector is given in \citet{K16}.

\begin{table}
 \caption{HERMES bandpasses}
 \label{bands}
 \begin{tabular}{@{}lc}
  \hline
  Channel & Wavelength range ($\hbox{\AA}$)\\
  \hline
  Blue & $4713 - 4903$\\
  Green & $5648 - 5873$\\
  Red & $6478 - 6737$\\
  IR & $7585 - 7887$\\
 \end{tabular}
\end{table}

Light is directed into HERMES from the 2dF fibre positioner \citep{LCT02}, which can place 400 magnetic ``buttons" carrying optical fibres across a circular field of view with a diameter of two degrees. It has two field plates with independent fibres, allowing one plate to be configured while the other is being used to observe. Eight of the buttons carry small fibre bundles that are used to maintain field alignment and telescope guiding, and the remaining 392 are single fibres that can be allocated to science targets and sky subtraction apertures. Twenty-five fibres are used as sky apertures, and a further 5-10 are typically unavailable for various engineering reasons. As a result, observed GALAH fields usually deliver spectra for around 360 science targets. 

\begin{figure*}
\resizebox{0.6\textwidth}{!}{\includegraphics{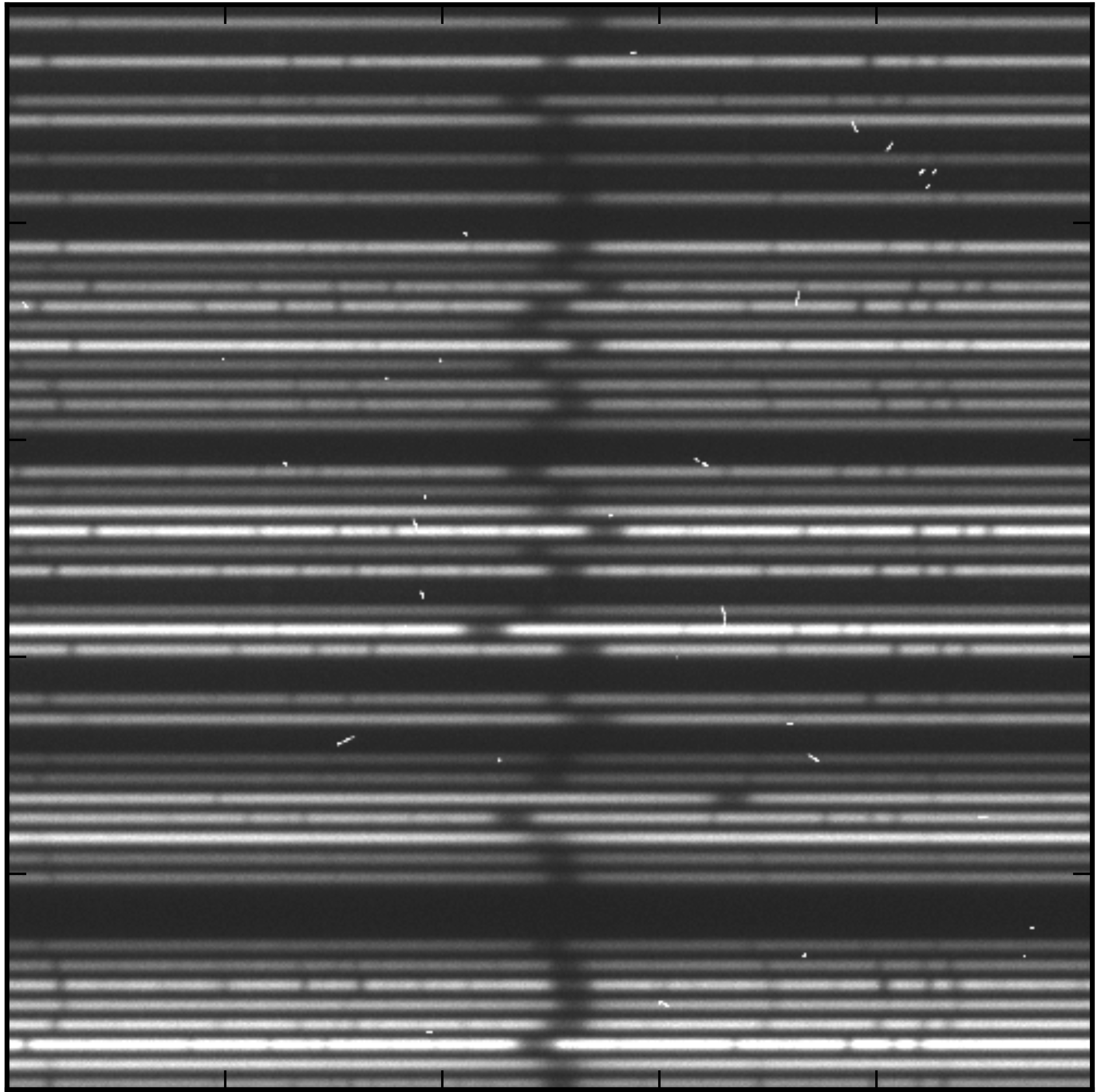}}
 \caption{A portion of a raw HERMES data frame from the red camera. The strong absorption feature near the centre of the image is H$\alpha$. The random scatter of anomalous pixels is due to unwanted charged particle hits.}
 \label{rawdata}
\end{figure*}

2dF is installed at the AAT prime focus, and the fibres run from the telescope top end to the HERMES enclosure on a lower floor of the AAT dome. The fibres are arranged in a pseudoslit at the spectrograph entrance, resulting in spectra on the detector with very similar wavelength range and dispersion. Figure \ref{rawdata} shows a zoomed-in region of typical raw data from the red channel: a 20-minute exposure in GALAH survey field 2235, observed on 9 August 2014. The dispersion direction is horizontal, and the fibres are separated vertically.

The point spread function (PSF) in HERMES varies across the spatial and spectral directions in all four cameras, with the smallest and most symmetric PSF in the centre of the detector and both ellipticity and tilt increasing toward the edges. This produces a small amount of crosstalk between spectra that are adjacent on the detector, which can be removed in data reduction \citep{K16}. In each wavelength channel, HERMES returns spectra with a SNR of $100$ per resolution element in one hour of exposure time in median (1\farcs5) seeing, for stars with a magnitude near $14$ in the corresponding filter (the exact limits are $B = 14.2$, $V = 13.8$, $R = 14.0$ and $I = 13.8$). Spectral response is fairly even across each channel, so (to first order) SNR is not a function of wavelength within each bandpass. This level of instrument efficiency (total throughput $\sim$10$\%$) was a design requirement for HERMES, to allow the GALAH survey to be completed in a reasonable amount of observing time.

There have been three known performance issues in the HERMES+2dF system: systematically higher throughput from targets on the northern half of both field plates, randomly distributed saturated points associated with long vertical readout streaks in three of the four cameras, and an inability to bring the red camera (which covers $6478$-$6757\hbox{\AA}$) entirely into focus. The North-South asymmetry in throughput, and smaller-scale throughput variations between fibres, are described in detail in \citet{SDS16}. It is not immediately clear what drives the North-South asymmetry, though it does not appear to be an issue in the fibres themselves. 

The vertical streaks are present in the blue, green and red cameras, but more common in the blue and green. They are believed to be caused by high-energy particles striking the detectors and freeing enough electrons to saturate a small number of pixels. These electrons are trapped low enough in the silicon layer that reading out the detector only partially flushes them out, so that the saturated pixels spawn a perfectly vertical feature that fades over the course of several exposures. The first time a given streak appears it runs away from the readout amplifier, and in all subsequent images it runs toward the readout amplifier. Laboratory testing by the Instrumentation group at the Australian Astronomical Observatory (AAO) has demonstrated that the high index-of-refraction glass in each camera's field flattening lens is likely to be the particle source. Most of the pixels affected by vertical streaks can be handled by the ordinary cosmic ray removal techniques employed in the GALAH data reduction process \citep{K16}.

In all four cameras, focus is achieved by adjusting the detector using actuators that move piston and tip the wavelength (``spectral") axis. The perpendicular (``spatial") axis was set during HERMES installation and commissioning, and is not movable through instrument software control. During HERMES downtime in June 2014, the detector in the red camera was tipped noticeably on its spatial axis, and engineering intervention was required. The spatial axis was returned to its original alignment and the piston was returned to its previous range. However, the new range of motion for the spectral axis was offset from the original range, and as a result it was no longer possible the actuators to move it sufficiently far to bring the red camera fully into focus. This issue was resolved on 10 August 2016 by AAO engineering staff, and the red camera can now be brought into focus as well as it could before June 2014. 

\begin{figure*}
\resizebox{0.6\textwidth}{!}{\includegraphics{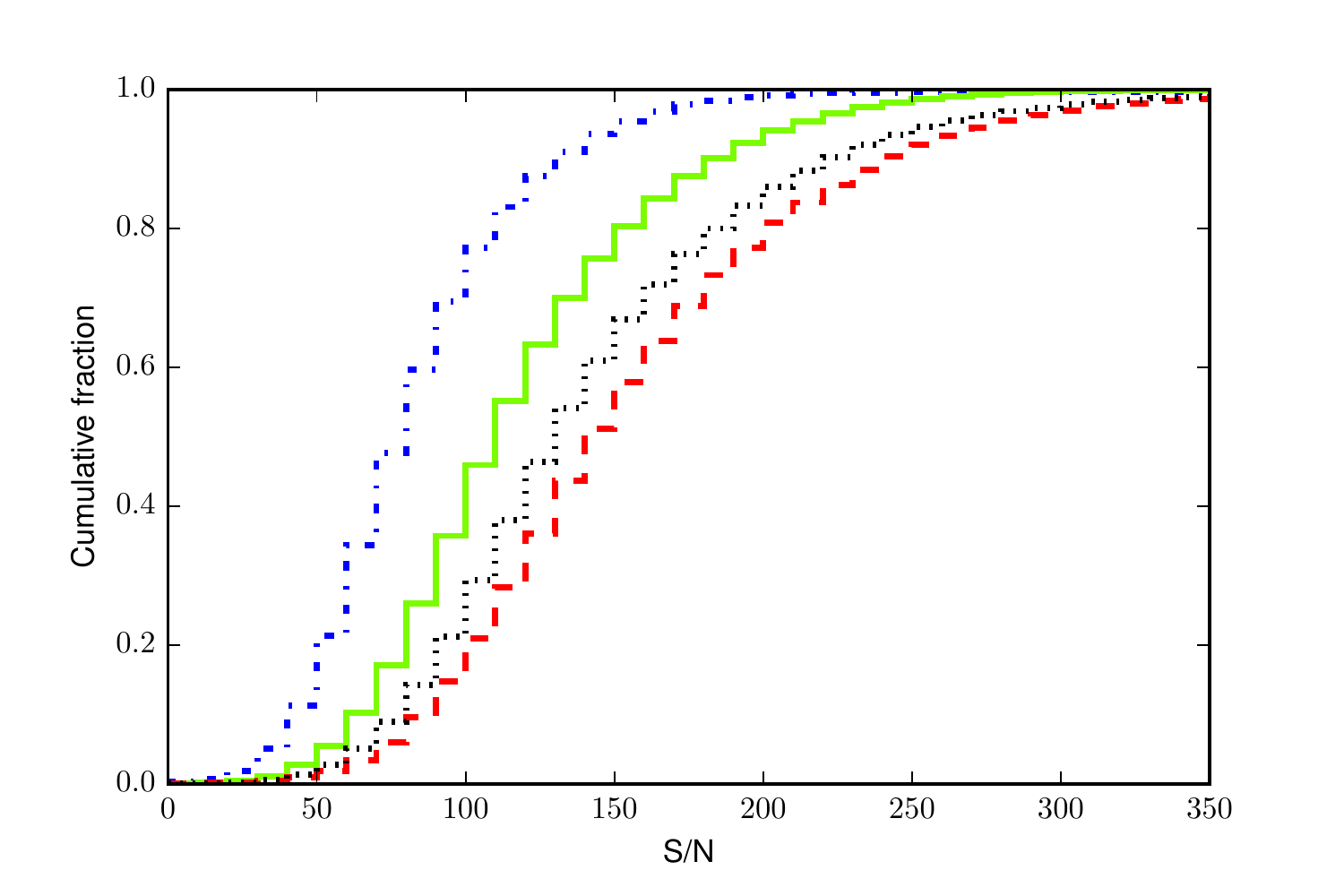}}
 \caption{Cumulative histograms of SNR per resolution element for the example data set, with the four HERMES channels drawn in different colours and line styles as described in the text.}
 \label{snrhist}
\end{figure*}

While HERMES does return spectra with SNR of 100 per resolution element in each camera for stars with apparent magnitudes near 14 in the appropriate Johnson-Cousins filter (as described above), only A-type stars, which are rare in the GALAH data set, have colours of zero and could have apparent magnitudes of 14.0 in $B$, $V$, $R$ and $I$ simultaneously. GALAH targets are selected based on a $V$ magnitude calculated from 2MASS $J$ and $K$ (as described in Section 3 below), and so we use the mean SNR per resolution element in the green channel spectrum as our figure of merit. As a result, the SNR for each star in each HERMES channel will be a function of its spectral energy distribution. Recent work within our team (\citealt{TCG15}; Ting et al., in prep) finds that increasing the resolution of chemical space by increasing the precision of abundance measurements is critical to large-scale chemical tagging, and that a SNR of at least 100 is required for abundance determinations as precise as 0.02 or 0.03 dex.

Figure \ref{snrhist} shows cumulative histograms of SNR per resolution element in each HERMES channel as reported by the data reduction software \citep{K16}. This figure shows data for 83026 unique science targets in a set of 256 regular survey fields with three 20-minute exposures (the typical GALAH observing pattern, as described in Section 4) observed between the start of the Pilot Survey (16 November 2013) and the end of AAT Semester 15B (30 January 2016). The blue dash-dot line represents the blue channel (in which $27\%$ of stars have SNR$>$100 per resolution element), the solid green line the green channel ($59\%$), the dashed red line the red channel ($82\%$) and the dotted black line the IR channel ($75\%$). In this ``example data set'' there are far more red stars than blue stars, and this will also be true of the full GALAH Survey. Therefore the SNR in the blue channel spectra will typically be lower than in the other three, while the SNR in the red and IR channel spectra are typically similar to each other and higher than in the green channel. 

\begin{figure*}
\resizebox{0.45\textwidth}{!}{\includegraphics{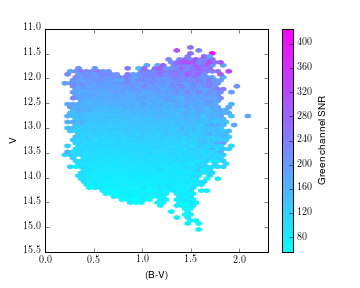}}\hspace{10pt}
\resizebox{0.45\textwidth}{!}{\includegraphics{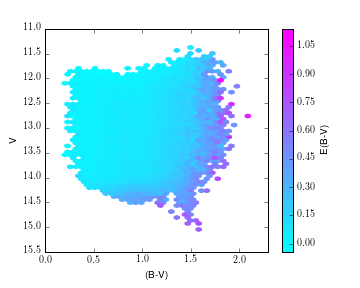}}\hspace{10pt}
 \caption{Apparent $(B-V), V$ colour-magnitude diagram for the example data set, binned into hexagons and colour-coded by mean green channel signal-to-noise ratio per resolution element in each bin (left panel) and mean $E(B-V)$ reddening in each bin (right panel).}
 \label{cmdsnr}
\end{figure*}

\begin{figure*}
\resizebox{0.6\textwidth}{!}{\includegraphics{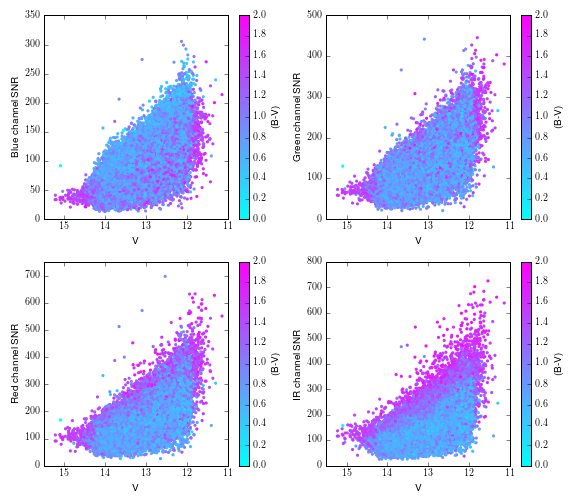}}
 \caption{Signal to noise ratio per resolution element for all four HERMES channels versus apparent $V$ magnitude for the example data set, colour-coded by $(B-V)$ colour. Bluer stars have higher blue channel SNR, but their SNR drops relative to the redder stars in the redder channels.}
 \label{vsnrbv}
\end{figure*}

Figure \ref{cmdsnr} shows the apparent (non-dereddened) Johnson-Cousins $(B-V), V$ colour-magnitude diagram for the example data set. These data have been binned into hexagons and colour-coded by mean green channel SNR in the bin (left panel) and by mean $E(B-V)$ reddening in the bin (right panel). Reddening is derived for each star as described in Section 9 below. There are three clear effects to be seen in this figure: first, that redder stars have a higher SNR at a fixed $V$ magnitude; second, that some of the  redder stars are simply more reddened rather than intrinsically redder; and finally, that the calculated $V_{\rm JK}$ magnitude we used for target selection does not always translate directly into the true Johnson-Cousins $V$ magnitude, but is moderated by stellar colour and by reddening. Intrinsically bluer stars must have brighter apparent $V$ magnitudes to be observed by GALAH than intrinsically redder stars. Figure \ref{vsnrbv} reinforces this point, showing SNR per resolution element in each HERMES camera in turn versus $V$ magnitude, colour-coded by $(B-V)$ colour. In the blue channel, the bluest stars have the highest SNR, but the SNR for these stars is clearly lower relative to the redder stars in the other three channels.

\section{Input Catalogue and Target Selection}
The GALAH input catalogue is the union of the 2MASS \citep{SCS06}, APASS \citep{MHF14} and UCAC4 \citep{ZFG13} catalogues, with selections for photometric quality and crowding. Because APASS  photometry was not available for all of our stars at the start of GALAH observing, we calculate a $V$ magnitude from 2MASS  $J$ and $K$ as follows: $V_{\rm JK} = K+2(J-K+0.14)+0.382e^{((J-K-0.2)/0.5)}$. All stars with apparent $V_{\rm JK}$ magnitude brighter than $14$ and Galactic latitude larger than five degrees are included in the input catalogue, provided that they have appropriate 2MASS quality flags\footnote{These flags are defined at http://www.ipac.caltech.edu/2mass/releases/allsky/doc/sec2\_2a.html} (Q=``A'', B=``1'', C=``0'', X=``0'', A=``0'', prox$\ge 6\arcsec$) and no brighter neighbours within a radius of $V_{\rm neighbour}=(130-(10\times V_{\rm neighbour}))$ arcseconds. This returns $5.99$ million stars.

To choose target stars from the input catalogue, we make no selection on colour or reddening, preferring a simple selection function that can be straightforwardly inverted to allow interpretation through Galactic models (e.g., \citealt{SBH14}). However, we do make some selections in support of survey science goals: declination, $\delta$, is limited to $-80<\delta <+10$ degrees, to maintain an airmass below 1.6 in all observations; Galactic latitude, b,  is restricted to $\mid b \mid > 10\degr$, to avoid significant and variable extinction closer to the plane; and the density of targets with $12 \le V_{\rm JK} \le 14$ must be at least $400$ per $\pi$ square degrees, to ensure efficient observations with 2dF. 

This more restricted set of $3.69$ million targets is then divided into $6545$ fixed ``configurations'' of $400$ stars each, to allow efficient use of the 2dF fibre positioner. These configurations use the full two-degree-diameter field of view in lower-density regions, and are more compact in denser regions, to allow a more efficient tiling. In particularly dense regions, multiple configurations can share a single field centre. The tiling strategy will be discussed in more depth in Sharma et al. (in prep). This survey sample is strongly focused on the thin and thick disk. Using the {\it Galaxia} software \citep{SBH11}, which simulates synthetic observation of the Milky Way using a Besan{\c c}on model \citep{RRD03} and Padova isochrones (\citealt{BB94}; \citealt{MG08}), we predict that $75\%$ of these stars belong to the thin disk, $24\%$ to the thick disk, $0.9\%$ to the bulge and $0.1\%$ to the halo.

Although we do not make any colour selections for survey science targets, we anticipate that our spectroscopic analysis will be most accurate and successful for stars with effective temperature between $4000$K and $7000$K. Stellar parameters and abundances will be more difficult to determine for stars outside that range: in hot stars, because of a lack of Fe and Ti lines in the HERMES wavelength ranges, and in cool stars, because of an overabundance of molecular features. We anticipate that the ongoing development of model atmospheres for cool stars (e.g., \citealt{A14}) will allow us to analyse those targets in the future. The lack of colour selection will also result a small minority of stars that are observed being found at extreme points of evolution for which our analyses will not work at all, e.g., T Tauri stars and white dwarfs.

\section{Observing Procedure}
As described above, HERMES meets the design requirement for a SNR of $100$ per resolution element when observing a target with $V_{\rm JK}=14$ for one hour in median seeing. This sets the basic unit of GALAH survey observing at one hour of integration time per field, with some adjustments as required depending on the observing conditions. The nominal GALAH observing procedure is to take three $20$-minute exposures for each configuration, with an additional $20$ minutes if the seeing is between $2\arcsec$ and $2\farcs5$ or an additional $60$ minutes if the seeing is between $2\farcs5$ and $3\arcsec$. We find that these adjustments to the exposure time are typically sufficient to raise the SNR to the required level (as discussed in \citealt{SAA15}). Typical overhead is $25\%$ of the on-target observing time for the standard $3\times20$ minute exposures. We spend $180$ seconds each for flat-field and ThXe arc exposures (taken directly before or after the science data), $71$ seconds per readout, and two to five minutes to slew the telescope, tumble the 2dF top end so that the other field plate is available for observing, and acquire the next science field. It takes $\sim$40~minutes to configure a full 2dF plate, which makes the total observation block time of $\sim$60~minutes well-suited to this efficient observing strategy, ensuring no observing time is lost due to plate reconfiguration.

There are a few special requirements for GALAH survey observing. To minimise the effects of chromatic variation and distortion in the 2dF corrector optics \citep{CFS08}, the change in airmass during a nominal $60$-minute GALAH observation will ideally be less than $0.05$. Because of the range of declination for GALAH targets, this translates into a strong preference that GALAH fields always be observed within $1.5$ hours of the meridian, unless there are no fields available at an appropriate hour angle. We also require that the field being observed is at least $30$ degrees from the Moon, since we are mainly observing in bright time, and that there are no bright Solar system planets within the field, since they have caused trouble in previous 2dF surveys. Stars from the input catalogue in the range $11<V_{\rm JK}<12$ are used as ``fiducial'' stars for field alignment and guiding during normal survey observations. We have developed software to select configurations for survey observations, and to keep track of which of the $6545$ survey configurations have been observed. This \textsc{ObsManager} software produces a list of configurations that meet the above criteria at a user-supplied date and time, produces the files used to configure the 2dF fibres, and tracks observational progress.

The 2dF configuration files produced by \textsc{ObsManager} contain lists of science targets, fiducial stars and sky positions, but they do not include specific allocations of targets to individual 2dF fibres. This information is added with the Configure program \citep{MSS06}, which uses a simulated annealing algorithm to assign the fibres to targets as efficiently as possible while also respecting the limits on where in the field each individual fibre can be placed, allocating a user-determined number of fibres to sky apertures, and maximising the number of guide fibres placed in the field of view.

Observations are made semi-classically. Although \textsc{ObsManager} could choose observable fields and produce setup files autonomously, the software controlling the 2dF fibre positioner and the HERMES spectrograph is not amenable to scripted operations, and the hardware occasionally needs human intervention. Decisions relating to variable seeing or weather also benefit from the intuition of an experienced observer. GALAH observations typically involve one or two astronomers from the science team, one of whom has significant experience observing with 2dF. These observers run \textsc{ObsManager}, select configurations to observe, configure 2dF, initiate all exposures, maintain raw data organisation and keep logs. In addition to observing at the AAT, observations are also routinely conducted remotely from the AAO offices in North Ryde and from remote observing facilities at Mt. Stromlo Observatory in Canberra and the International Centre for Radio Astronomy Research in Perth.

\section{GALAH Pilot Survey}
The GALAH Pilot Survey, which ran from 16 November 2013 until 19 January 2014, was a joint science verification and early science program, concurrent with HERMES commissioning. There were four main projects in the Pilot Survey: {\it Gaia} benchmark stars, thin/thick disk normalisation, star clusters, and asteroseismic targets observed by the CoRoT satellite.  These projects covered a wide range of possible uses for HERMES, while allowing the commissioning and science verification teams to test critical functions of both the instrument and the GALAH software. The data set and goals for each of these projects are described below; results will be published separately as each project progresses. Because of the restricted range in target right ascension, the observing procedure was not as strict for the Pilot Survey as for the main survey, and fields were observed at hour angles between $-$01h:45m and $+$06h:30m (though this extreme case was for a circumpolar field). 

\subsection{{\it Gaia} benchmark stars}
We have observed 26 of the 34 stars designated as benchmark stars for the {\it Gaia} mission \citep{HJG15}. Since these stars are all quite bright, we observed them with a single 2dF fibre rather than as part of regular survey configurations. Exposure times were short, typically less than 120 seconds, such that telescope tracking was sufficient to maintain the alignment of the star on the fibre. These stars have weakly-model-dependent measurements of their stellar parameters based on angular diameter, bolometric flux, and parallax, which can be used to test the accuracy of spectroscopic stellar parameter determinations. They are also an excellent (if small) data set for cross-survey comparison and calibration, since they are well distributed across parameter space and evolutionary state, and across the sky. GALAH stellar parameters and metallicities for {\it Gaia} benchmark stars will be discussed in a future paper on the data analysis pipeline.

\subsection{Thin/thick disk normalisation}
The largest amount of observing time in the Pilot Survey was spent on a program to investigate the normalisation (that is, the ratio of thin to thick disk stars in the midplane) and rotational lag between the Galactic thin and thick disks. A clear chemical separation in the ${\rm [}\alpha /{\rm Fe]} - {\rm [Fe/H]}$ abundance plane can be made between these two populations (see, e.g., \citealt{AS12}; \citealt{BFO14}; \citealt{HBH15}). With HERMES spectra, we can study the overall ${\rm [}\alpha /{\rm Fe]} - {\rm [Fe/H]}$ plane and the behaviour of individual alpha elements in the thin versus the thick disk, since they do not all have the same nucleosynthetic origins.

The intended observational targets of this project were red giant branch stars $\sim$3.3~kpc from the Sun. They were selected from the 2MASS catalogue with $(J-K) > 0.45$ for $10 < K < 12.2$ and $(J-K) > -0.1$ for $9 < K < 10$ and the same quality flags as the GALAH input catalogue. This program took observations of 9847 stars in 29 fields with Galactic longitude $l \sim$270$\degr$ and latitude $b$ of $-16\degr{\rm,}$ $-22\degr{\rm,}$ $-28\degr{\rm,}$ $-35\degr$ and $-42\degr$. Because targets were chosen based on photometry, there was contamination by foreground dwarfs. Separating dwarfs and giants at a surface gravity of log(g)$=$3.8, the contamination was typically $36\%$, rising for stars further from the plane. This is somewhat lower than the dwarf/giant ratio we find in regular GALAH survey observations in the disk, indicating that the colour selection was helpful in isolating giants. The results of this project will be presented in Duong et al. (in prep).

\subsection{Globular and open clusters}
Globular and open star clusters provide important anchor points for large stellar surveys like GALAH (e.g., \citealt{SLB11}; \citealt{AZS15}). We use stars in globular and open clusters to confirm that our analysis pipelines are returning reasonable and consistent values for stellar parameters and abundances, to cross-calibrate with other large survey projects, and as benchmarks for chemical tagging methods.

The Pilot Survey included targeted observations of stars in four globular clusters (NGC 288, NGC 362, NGC 1851 and 47 Tucanae) and the open cluster M67. These clusters were selected to provide a broad coverage of metallicity, and for observability with HERMES during the Pilot Survey: right ascension, $\alpha$, in the range 0h $< \alpha <$ 9h and distance modulus, $(m-M)_V$, less than $15.5$. These observations were taken differently from normal GALAH survey observations, with the apparent magnitude range extended as faint as $V = 16$ and the total exposure times extended as long as 6 hours per field for the more distant clusters to capture as many stars as possible from the red giant branch, red clump and horizontal branch. The globular clusters $\omega$ Centauri and NGC 7099 were also specifically targeted after the end of the Pilot Survey, to provide additional well-studied anchors for our analysis. Similar to the Pilot Survey clusters, the apparent magnitude range extended to $V = 17$ and the exposure times were as long as 6.3 hours.

\begin{figure*}
\resizebox{0.6\textwidth}{!}{\includegraphics{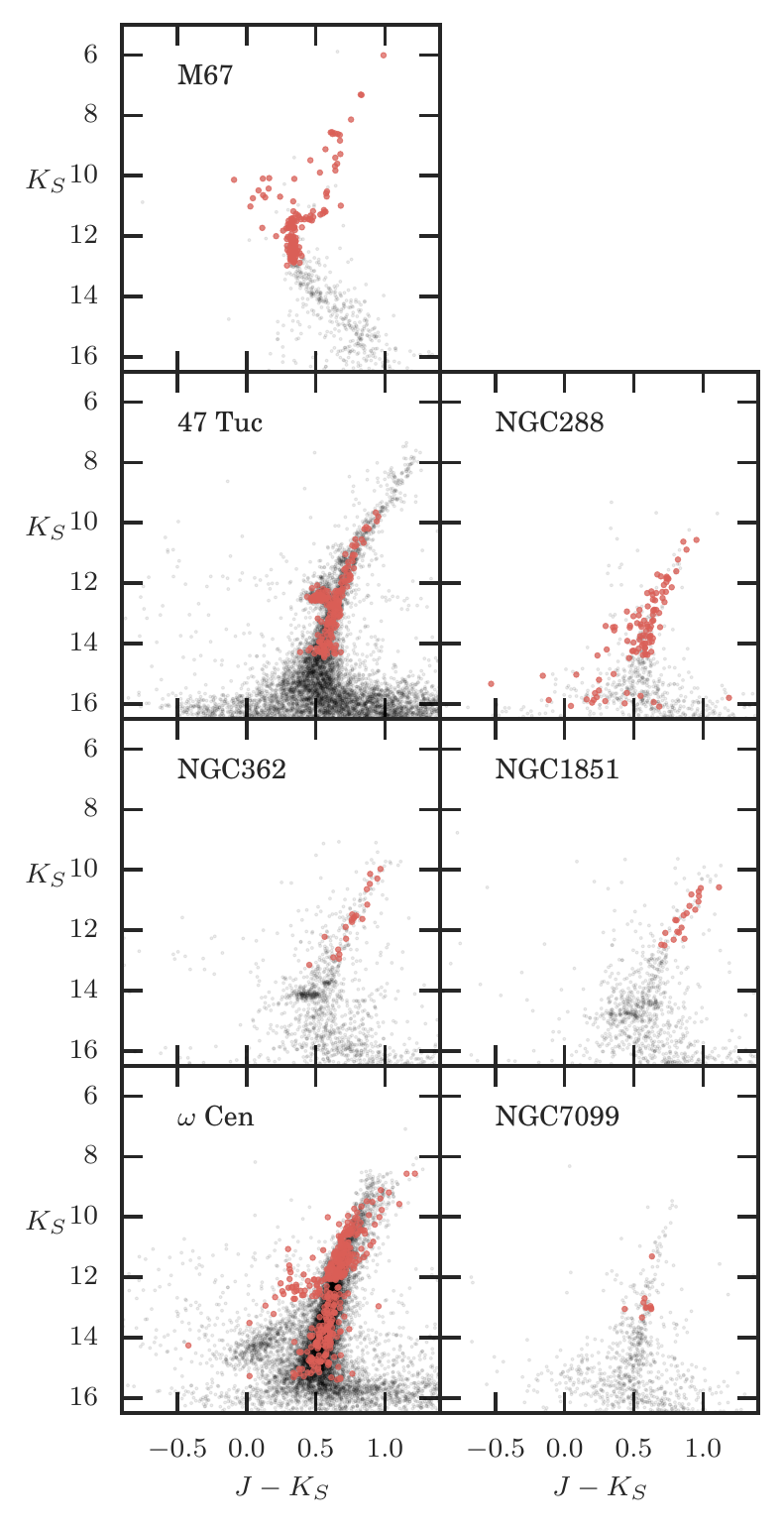}}
 \caption{Near-infrared colour-magnitude diagrams for the seven open and globular clusters observed intentionally. Red points are the cluster members, and smaller grey points are all stars in the 2MASS Point Source Catalogue within 10$\arcmin$ of cluster centre.}
 \label{cmds}
\end{figure*}

Targets in the Pilot Survey clusters were chosen from cluster members identified in previous studies (Stetson, priv. comm.; Da Costa, priv. comm.; \citealt{CBG09}; \citealt{YG09}; \citealt{SC13}; \citealt{MM14}; \citealt{NMZ15}; \citealt{D16}). Targets in $\omega$ Cen were taken from \citet{BPB09}, and in NGC 7099, targets were taken from \citet{D16}. We were only able to observe between $10$ and $173$ cluster members in any single configuration, given the magnitude limits and the limitations of the fibre positioner (2dF fibres cannot be placed closer together on the sky than $30\arcsec$). All together, we observed between $10$ and $394$ stars total per cluster, typically in the outer regions. Table \ref{clusters} lists coordinates, distance moduli, metallicity (taken from \citealt{H96}, 2010 edition for the globular clusters and \citealt{HS14} for M67), number of stars observed, $V$ magnitude range, exposure time, and dates of observation for all of the globular and open clusters observed in this targeted fashion. Figure \ref{cmds} shows colour-magnitude diagrams for all of these clusters, with stars observed by GALAH highlighted as red circles and stars from the 2MASS Point Source Catalogue within $10\arcmin$ of cluster centre shown as smaller grey circles. 

\begin{table}
 \tabcolsep=0.11cm
 \caption{Data for globular and open clusters observed intentionally by GALAH}
 \label{clusters}
 \scalebox{0.9}{
 \begin{tabular}{@{}lllllrlll}
  \hline
  Cluster & $\alpha$ & $\delta$ & (m-M)$_{V}$ & [Fe/H] & N$_{\rm stars}$ & $V$ & $t_{\rm exp}$ (s) & Obs. dates\\
  \hline
  M67 & 08:51:18 & $+$11:48:00 & 9.97 & 0.0 & 140 & 8.8$-$14.0 & 3600$-$7200 & 2013 17 Dec, 2014 09 Feb\\
  \hline
  47 Tuc & 00:24:05.67 & -72:04:52.6 & 13.37 & -0.72 & 156 & 12.1$-$16.0 & 4800$-$21480 & 2013 20 Nov, 23 Nov, 19 Dec,\\
  &&&&&&&& 20 Dec, 2014 11 Jan, 13 Jan\\
  NGC 288 & 00:52:45.24 & -26:34:57.4 & 14.84 & -1.32 & 104 & 13.0$-$16.0 & 7200$-$19200 & 2013 18 Nov, 20 Nov,\\ 
  &&&&&&&& 2014 14 Jan, 15 Jan, 16 Jan\\
  NGC 362 & 01:03:14.26 & -70:50:55.6 & 14.83 & -1.26 & 21 & 12.7$-$14.7 & 7200 & 2013 18 Nov, 23 Nov\\
  NGC 1851 & 05:14:06.76 & -40:02:47.6 & 15.47 & -1.18 & 20 & 13.3$-$14.6 & 7200 & 2014 15 Jan, 17 Jan\\
  $\omega$ Cen & 13:26:47.24 & -46:28:46.5 & 13.94 & -1.53 & 394 & 12.0$-$17.0 & 3200$-$22800 & 2014 03 Mar, 04 Mar,\\ 
  &&&&&&&& 05 Mar, 07 Mar\\
  NGC 7099 & 21:40:22.12 & -23:10:47.5 & 14.64 & -2.27 & 10 & 13.1$-$15.0 & 7200 & 2015 01 Sep\\
  \hline
 \end{tabular}}
\end{table}

\begin{figure} 
\resizebox{0.45\textwidth}{!}{\includegraphics{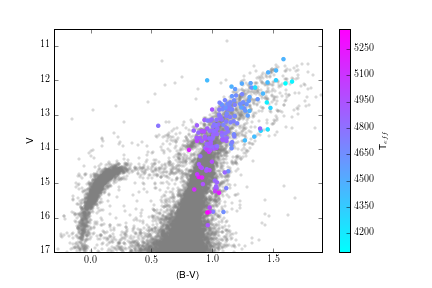} }\hspace{10pt} \resizebox{0.45\textwidth}{!}{\includegraphics{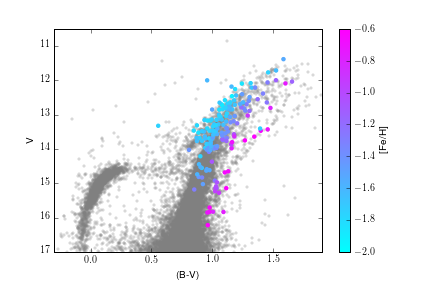} } 
\caption{The (V,B-V) colour-magnitude diagram for $\omega$ Centauri, colour-coded by GALAH effective temperature (left panel) and metallicity (right panel). All stars within $10\arcmin$ of the cluster centre, with membership probability from \citet{BBP10} above 0.9, are shown as smaller grey circles. Both T$_{eff}$ and [Fe/H] behave as expected, both in terms of range and trends. 
\label{wcen_cmds}}
\end{figure}

Figure \ref{wcen_cmds} shows the $(V,B-V)$ colour-magnitude diagram for $\omega$ Centauri. All stars within $10\arcmin$ with a membership probability above 0.9 are shown as small grey circles, and stars observed by GALAH are highlighted as larger coloured circles. In the left panel they are colour-coded by our derived effective temperature, and in the right panel they are colour-coded by our derived metallicity. The optical photometry is taken from \citet{BBP10}, which we also used for spectroscopic target selection. Our derived $T_{\rm eff}$  follows exepcted trends, and our derived [Fe/H] values show an overall similarity to the complex morphology described in \citet{JP10}, with the reddest giant branch being the most metal-rich.

\begin{figure*}
\resizebox{0.6\textwidth}{!}{\includegraphics{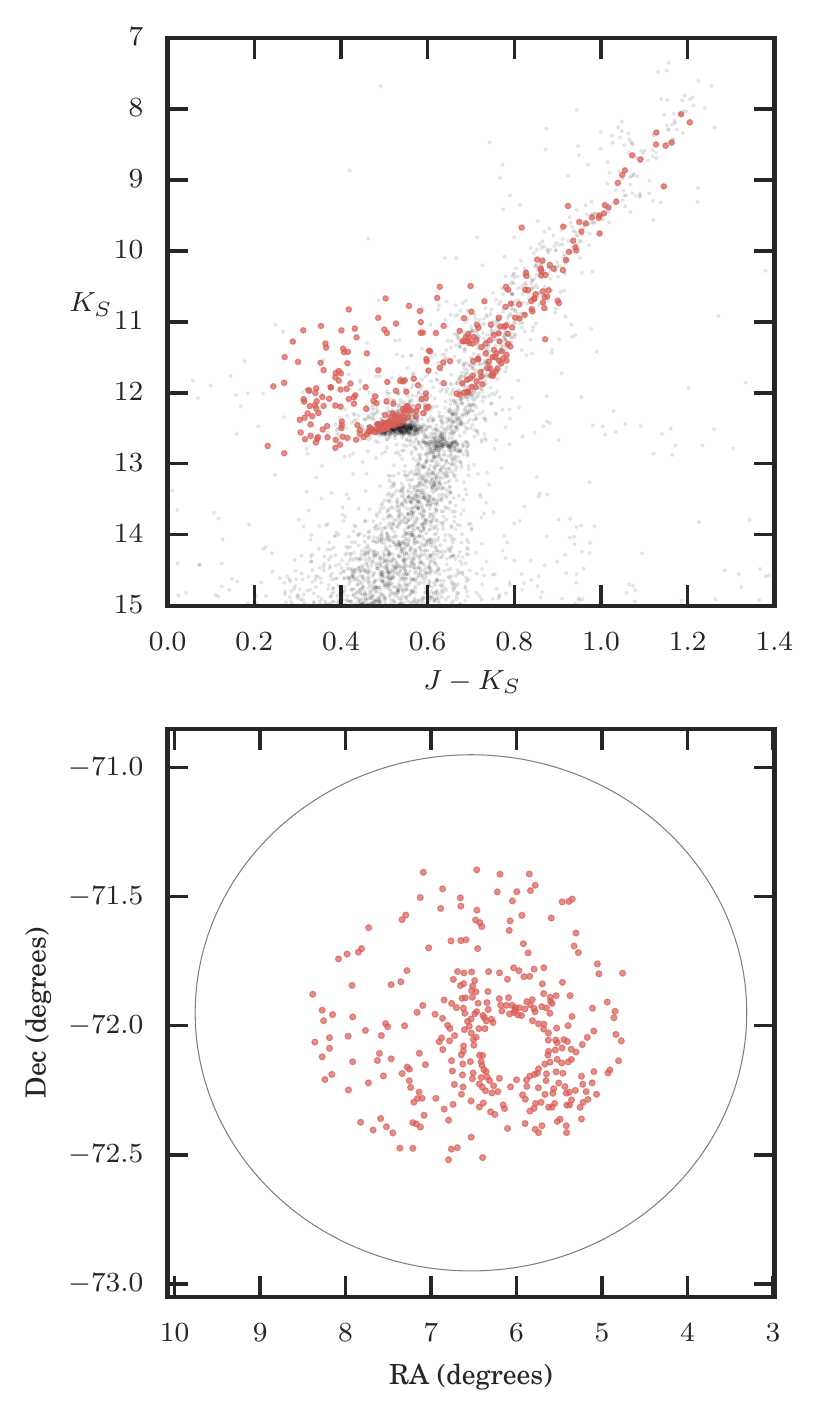}}
 \caption{Near-infrared colour-magnitude diagram for the stars in GALAH survey field 51. Red points are all stars observed in the field, and smaller grey points are all stars in the 2MASS Point Source Catalogue within 10$\arcmin$ of the centre of 47 Tuc, similar to Fig. \ref{cmds}. Stars that are likely cluster members based on their photometry can be confirmed using radial velocities, stellar parameters and abundances determined from the spectra.}
 \label{scmd}
\end{figure*}

In addition to the stars observed intentionally during the Pilot Survey, a number of cluster members have been observed serendipitously in GALAH survey fields. The upper panel of Figure \ref{scmd} shows a colour-magnitude diagram for the 318 stars observed in survey field 51 (red circles), and the lower panel shows the spatial distribution of targets for that field, with a circle marking the field of view of 2dF. The concentration of targets near 47 Tuc is clear in the south-western quadrant of the field, and in the colour-magnitude plane the cluster red giant branch can be seen mixed together with the broader distribution of field stars. 2MASS photometry for all stars within $10\arcmin$ of the centre of 47 Tuc is also shown as small grey circles to guide the eye. Membership for serendipitously observed cluster stars can be verified with radial velocity and proper motion. In addition to these serendipitously observed 47 Tuc stars, we have identified stars belonging to NGC 362, M67, NGC 2516, NGC 2243, NGC 6362 and the Pleiades within regular survey fields, and we anticipate that future survey observations will provide serendipitous observations of many more cluster members and extratidal stars associated with star clusters (e.g., \citealt{NMZ15}). 

\subsection{CoRoT targets}
The intersection of asteroseismic and spectroscopic data opens a number of new possibilities for Galactic archaeology. Prior to 2015, the only large-scale asteroseismic mission with targets that could be observed from the Southern hemisphere was CoRoT \citep{ABB09}. CoRoT observed one large region in the direction of the Galactic centre and one toward the anticentre, both at declination near zero, potentially providing common targets for GALAH and other ongoing Galactic archaeology surveys.  We observed $2218$ stars in six configurations in the CoRoT anticentre fields LRa01, LRa05 and LRa07 as part of the pilot survey, with a simple $12 < $V$_{\rm JK} < 14$ magnitude selection to match the target selection in the main GALAH survey. Of these, 1526 have successfully been processed through the GALAH analysis pipeline (as described in Section 9 below). We find that 737 of these stars have surface gravity below 3.5, making them ideal for asteroseismic interpretation. Results from this project will be presented in Anguiano et al. (in prep).

\section{The K2-HERMES program}
Following the failure of a second reaction wheel onboard the Kepler spacecraft in 2013, the extremely precise pointing that enabled its photometric monitoring work in a single Northern field was lost, and the spacecraft was re-purposed to an observational program along the ecliptic that enabled it to regain its fine pointing precision using only the two remaining reaction wheels and regular thruster firings. A series of 80-day observing ``campaigns'' along the ecliptic were laid out, and the community was invited to propose targets for this ``K2'' mission. The data acquired by K2 are useful both for asteroseismology and for transiting exoplanet surveys.

There is tremendous potential in combining asteroseismic information with stellar surface temperature and composition data from photometry and spectroscopy (e.g., \citealt{MCM13}; \citealt{CSA16}), bringing a new level of precision to determinations of stellar age, mass, radius and surface gravity for red giant stars. Combined asteroseismic and spectroscopic data sets allow us to study stellar populations on a Galactic scale (e.g., \citealt{SHS15}; \citealt{SSB16}; \citealt{MFR16}). They also make asteroseismic data useful across a broader range of stellar metallicity by providing crucial calibrations for the scaling relations used to interpret stellar oscillations \citep{EE14}. In addition, stellar parameters obtained from spectroscopy are critical for determining the sizes of the transiting planet candidates detected by K2, because uncertainties in planet size are dominated by uncertainties in the stellar radius (e.g., \citealt{K14}; \citealt{WWO05}).  Photometrically derived stellar radii such as those from the Kepler Input Catalogue have been shown to have uncertainties of up to $\sim 40\%$ for Solar-type stars (e.g., \citealt{VC11}; \citealt{EH13}; \citealt{BS14}), and similar uncertainties apply for the majority of K2 targets that have been classified in the Ecliptic Plane Input Catalogue \citep{HB16}. When high-resolution, high S/N spectra are used in combination with transit measurements, planetary radii can be determined to precisions of 10$-$15\% (e.g., \citealt{SAD15}; \citealt{WRI16}). 

Galactic archaeology target proposals for K2, spearheaded by GALAH team members S. Sharma and D. Stello, have been very successful, with typically 5000 targets in each K2 observing campaign (though not all of these targets will turn out to be giants). Through a separate K2-HERMES observing program at the AAT (AAT 15A/03, 15B/01, PI Wittenmyer; AAT 15B/03, 16A/22, PI Sharma), many of these targets as well as potential exoplanet hosts, are now being observed using similar procedures as for GALAH, enabling beneficial collaboration between GALAH and the K2-HERMES program. In exchange for data reduction and processing with the GALAH pipeline, K2-HERMES data are incorporated into the GALAH Survey database. K2-HERMES stars with asteroseismically derivable parameters will be quite helpful in the testing and refinement of the GALAH analysis pipeline. 

The K2 field of view consists of nineteen separate square ``CCD modules'' covering four square degrees each, and the K2-HERMES fields lie at the centre of each CCD module's field of view. The spectroscopic target selection is different in each K2-HERMES field based on its Galactic coordinates, targeting different stellar populations in different lines of sight. K2-HERMES configurations typically cover a wider range in apparent magnitude than GALAH survey fields. As with the survey fields, K2 fields are observed within 90 minutes of the meridian. As of 30 January 2016, the K2-HERMES program has observed 31,365 stars.

\section{Observational Progress}

\begin{figure*}
\resizebox{1.0\textwidth}{!}{\includegraphics{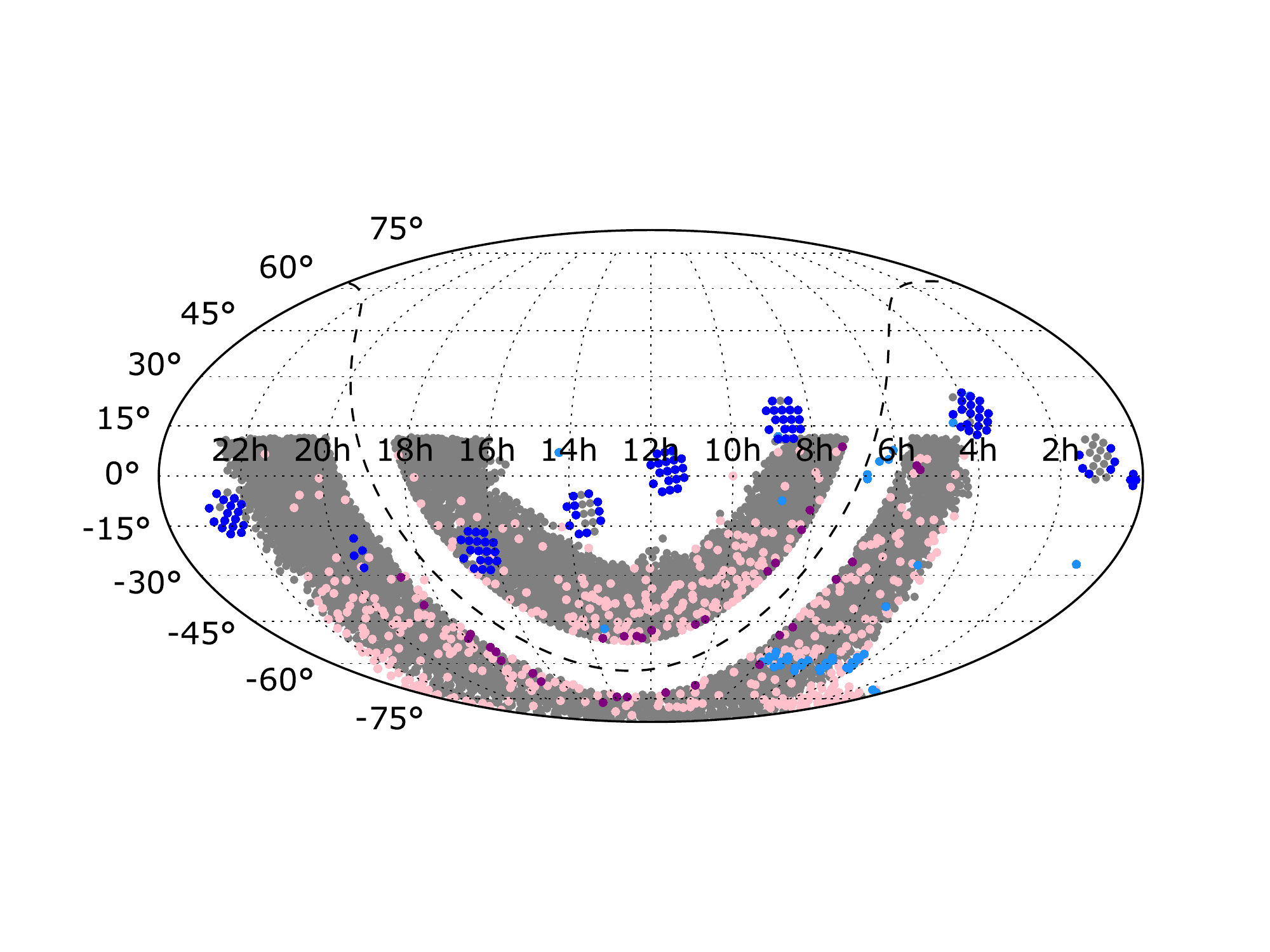}}
 \caption{Map of GALAH Survey progress through 30 Jan 2016. Grey circles are unobserved survey fields, pink are regular survey fields that have been observed, cyan are fields observed during the Pilot Survey, blue are fields observed by the K2-HERMES program, and purple are fields observed for the targeted {\it Tycho-2} bright star subproject.}
 \label{progressmap}
\end{figure*}

With a large allocation of time (26 nights for the Pilot Survey in Semester 13B and 70 nights per year for the full survey starting in Semester 14A) and a highly multiplexed spectrograph, GALAH observing progress has moved quickly despite poorer-than-average weather. Our data rate is 4.2 stars per minute spent on-sky, yielding roughly 50,000 stars per semester. Figure \ref{progressmap} is an equatorial-projection map of observing progress through 30 January 2016. In this map, grey circles are unobserved survey fields, pink are regular survey fields that have been observed, purple are fields observed for the Tycho-2 bright star subproject (described in Section 9 below), blue are fields observed by the K2-HERMES program, and cyan are fields observed during the Pilot Survey. The number of observable regular survey fields varies strongly with right ascension, with all survey fields in the range 23h$\leq \alpha \leq$4h already observed once. 

\begin{figure*}
\resizebox{0.6\textwidth}{!}{\includegraphics{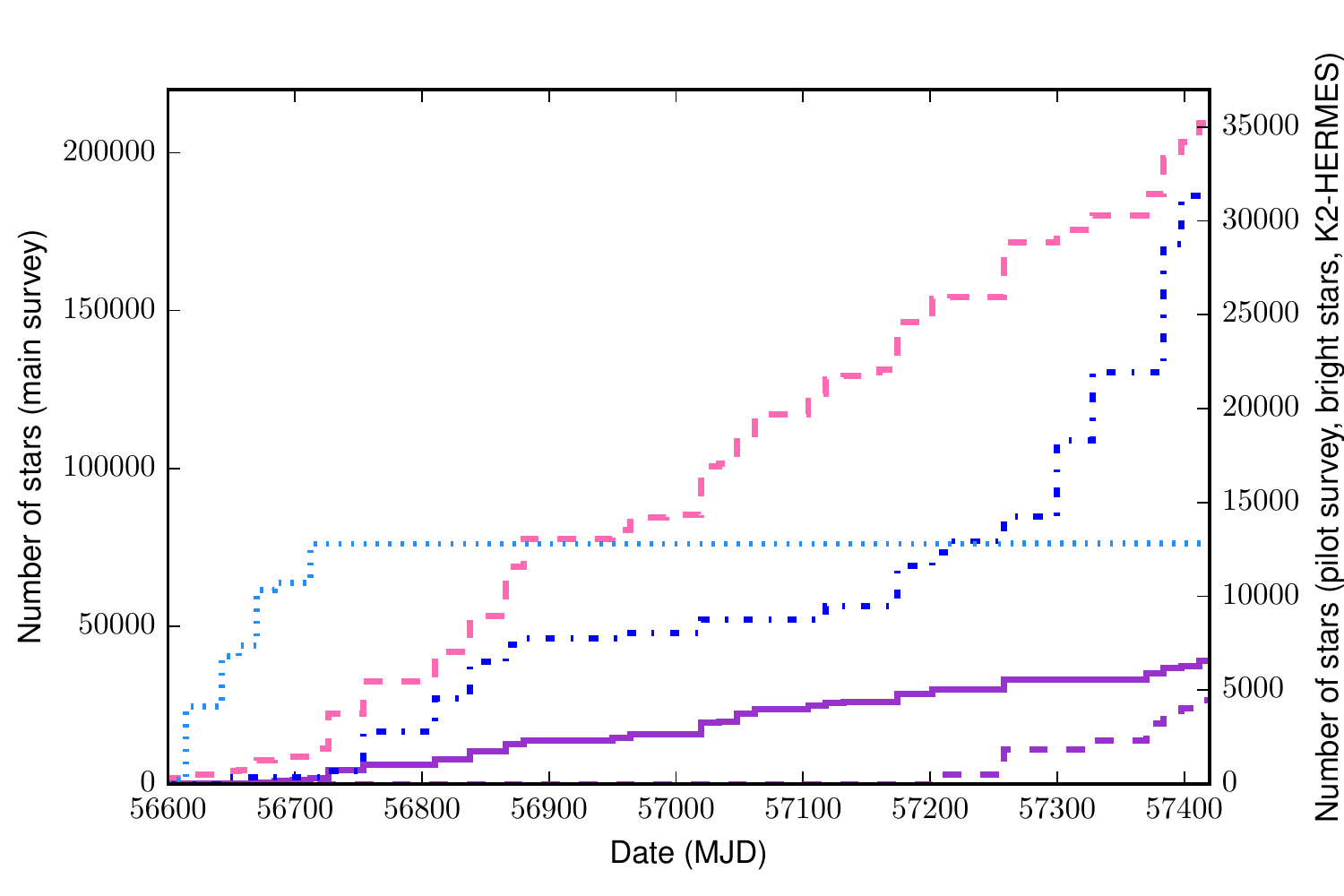}}
 \caption{Cumulative histogram of the number of stars observed versus MJD. with colours denoting different survey subsets (dashed pink: main survey; dash-dot dark blue: K2-HERMES; dotted cyan: pilot survey; dashed purple: targeted {\it Tycho-2} stars; solid purple: serendipitous {\it Tycho-2} stars. The vertical axis on the left is for the main survey, and the vertical axis on the right is for the other projects.}
 \label{numberhist}
\end{figure*}

Figure \ref{numberhist} is a cumulative histogram of the number of stars observed from the start of the Pilot Survey through 30 January 2016 in each of those programs, in fortnightly bins. Since there are significantly more stars that have been observed for the regular survey than the other programs, the y axis on the left is for the regular survey, and the y axis on the right is for the Pilot Survey, K2-HERMES stars and {\it Tycho-2} bright stars. The number of regular survey stars is shown with a pink dashed line, the number of Pilot Survey stars is shown with a cyan dotted line, the number of K2-HERMES stars is shown with a blue dash-dot line, and the number of {\it Tycho-2} bright stars is shown with a purple solid line ({\it Tycho-2} stars observed serendipitously during other observations) and a purple dashed line (targeted bright-star fields). The Pilot Survey starts first and runs for a few months, the regular survey and the K2-HERMES program start shortly before the Pilot Survey ends, and the targeted {\it Tycho-2} bright star observations begin later. Through 30 January 2016 we have observed 209,345 stars in the main survey, 12,910 stars in the various Pilot Survey programs, and 11034 {\it Tycho-2} stars (4448 targeted and 6586 observed serendipitously), and an additional 31,365 stars have been observed by the K2-HERMES program. 

\section{Survey Synergies}
The wide sky coverage of the GALAH Survey provides significant overlap with several other large-scale surveys. This creates important synergies, allowing us to link our thorough local sample with the astrometric measurements from the {\it Gaia} mission, the pencil-beam {\it in situ} halo samples of the {\it Gaia}-ESO and APOGEE surveys, the thorough {\it Gaia}-ESO coverage of open clusters, the low-latitude disk sample from APOGEE and a significant fraction of the very large sample of the RAVE survey.

The {\it Gaia} satellite (\citealt{P12}; \citealt{LP96}), launched in late 2013, is collecting high-precision astrometry and photometry for stars with apparent magnitudes $5.7 < V < 20$ as well as moderate-resolution spectroscopy near the near-infrared calcium triplet and low-resolution spectrophotometry for stars down to $V = 17$. {\it Gaia}'s full catalogues will be revolutionary for our understanding of the phase-space structure of the Galaxy, as well as providing spectrophotometry and basic stellar parameters for as many as one billion stars. Perhaps the most important synergy we have is with {\it Gaia}, as the entire GALAH input catalogue is within the brightest $1\%$ of {\it Gaia} targets, which will have parallax uncertainties less than 10 $\mu$as and proper motion uncertainties of less than 10 $\mu$as per year, corresponding to $1\%$ distance errors and $0.7$ kms$^{-1}$ velocity errors at $15$ kpc. Coupling the unprecedented abundance detail of GALAH with the 6-dimensional phase-space positions and velocities that can only be measured by {\it Gaia} will allow us to identify chemically homogeneous groups of stars that also match in age and orbital properties, revealing the process of star formation and chemical evolution in the Galaxy.

The {\it Gaia}-ESO Survey (GES; \citealt{GRA12}; \citealt{RG13}) is another ongoing Galactic archaeology survey project, using the GIRAFFE spectrograph \citep{PAA00} at the Very Large Telescope at the European Southern Observatory in Chile to collect high-resolution ($R\sim 26,000$) spectra for $100,000$ stars, primarily in the halo and in star clusters. A smaller sample of brighter stars is also being observed at higher resolution ($R\sim 47,000$) with the UVES spectrograph \citep{DDK00}. Because of the wavelength regions accessible to the multiobject modes of GIRAFFE and UVES, and the resolution of the spectra they produce, GES can determine stellar parameters and abundances for as many as 15 and 34 elements, respectively, per star.

The APOGEE Survey and its followup project, APOGEE-2, are components of the third and fourth iterations, respectively, of the Sloan Digital Sky Survey. APOGEE targeted over 150,000 red giants across the disk, bulge and halo, with a regular grid pattern across the sky. APOGEE observations were carried out from Apache Point Observatory in the United States, and APOGEE-2 will continue observations from the same observatory and begin a Southern observing campaign using a duplicate spectrograph at the Ir{\'e}n{\'e}e du Pont telescope at Las Campanas Observatory in Chile. This survey's unique advantage is that its wavelength coverage is entirely in the infrared ($1.51\mu$m$ - 1.70\mu$m), reducing the line-of-sight extinction and allowing observations of stars much closer to the Galactic plane, including stars on the far side of the bulge. APOGEE spectra have a resolution of $R\sim$22,000; the most recent data release\footnote{http://www.sdss.org/dr13/} includes abundances for up to 21 elements per star \citep{GP16}.

The large telescope and excellent site at Paranal allow GES to observe fainter stars than those targeted by GALAH, enabling the capture of a larger sample in the bulge and the {\it in situ} halo, albeit with a pencil-beam distribution. APOGEE also typically targets stars at larger distances than GALAH does, but perhaps more importantly it includes a significant sample of stars in and near the plane of the Galaxy. The combination of GALAH, GES and APOGEE data will enable science that cannot be done by any one of the surveys alone. Potential examples are studies of radial and vertical trends in the thin disk that use the local GALAH sample as an anchor and the more distant APOGEE and GES samples as probes. It will be critical to bring the abundance results of these different projects onto the same scale to allow this type of cross-survey study. 

There is some observational overlap designed in to GALAH, GES and APOGEE, despite their different selection functions, to facilitate this cross-calibration. The Southern extension of APOGEE-2 will make cross-calibration between GALAH and APOGEE much more straightforward, and APOGEE-2 observations are planned to provide a set of stars that comprehensively cover the parameter space of GALAH and APOGEE stars. We have already identified a serendipitous survey overlap of 185 stars with GES, evenly split between UVES and GIRAFFE observations, and a serendipitous overlap of 664 stars with APOGEE, mainly in the K2 ecliptic campaign fields and CoRoT Galactic anticentre regions. The data-driven approach of The Cannon \citep{NHR15a} will be central to the cross-calibration effort (e.g., \citealt{H16}), and its capabilities in this area have already been demonstrated in \citet{NHR16}.

The Radial Velocity Experiment (RAVE, \citealt{KGS13}) survey is an important precursor to the current generation of Galactic archaeology surveys. RAVE took R$\sim$7500~spectra in a small wavelength range near the calcium triplet for nearly 500,000 stars with $9 <$I$ < 12$. While the original plan was to determine radial velocities and basic stellar parameters from these data, the RAVE team has shown that they can also derive several elemental abundances and probabilistic distances (\citealt{BB14}; \citealt{BSW11}; \citealt{KRB11}; \citealt{ZMB10}). The development of automated spectrum analysis pipelines has benefited from the work of the RAVE team and their goal of maximising the amount of information to be derived from their spectra. A large fraction of the RAVE sample falls into GALAH's input catalogue, since both are Southern-sky surveys primarily using apparent magnitude for target selection. There are 9388 RAVE stars in the data set considered in this paper. Having GALAH observations of a large number of RAVE stars provides an extremely large comparison set for GALAH radial velocities, and will enable detailed followup and extension of important RAVE studies of Galactic dynamics and structure (e.g., \citealt{WSB13}; \citealt{ABB12}, \citealt{SFB12}; \citealt{RFW11}).

SkyMapper \citep{KSB07} is an Australian synoptic survey project imaging the Southern sky in 6 photometric bands. Its particular advantage is the inclusion of a Str{\" o}mgren-like $u$ filter that captures the Balmer jump and a narrow $v$ filter that spans the Ca II H and K lines, similar to the DDO38 filter. Colour indices including these filters can be constructed to be quite sensitive to either surface gravity or metallicity (e.g., \citealt{KBF14}; \citealt{HAC14}). SkyMapper photometry will be a useful tool for a number of GALAH science goals, including the identification of very metal-poor stars, confirmation of star cluster membership, and the study of interstellar reddening through comparison of stellar effective temperatures derived photometrically and spectroscopically. We have already identified roughly 60,000 stars in common between GALAH and the SkyMapper Early Data Release, which includes objects from their ``short survey" of relatively bright targets. We expect that ultimately all GALAH stars will be in the SkyMapper catalogue.

\section{{\it Tycho-2} stars and {\it Gaia} DR1}
The {\it Tycho-2} catalogue \citep{HFM00} contains positions and magnitudes for 2.5 million stars.  Although the full precision of the astrometric solution for the full {\it Gaia} dataset can only be reached with several years of data, combining the {\it Tycho-2} catalogue with the first year of {\it Gaia} data (at an epoch 24 years later) allows a precise solution for positions ($\sigma \le 0.75$ mas), parallaxes ($\sigma \le 0.64$ mas) and proper motions ($\sigma \le 3.19$ mas yr$^{-1}$) for all the {\it Tycho-2} stars (the ``Tycho-Gaia Astrometric Solution'', TGAS), as described in \citet{MLH14} and \citet{MLH15}.

In anticipation of the first {\it Gaia} data release and the TGAS work, GALAH has prioritised observations of {\it Tycho-2} stars, generating 330 special configurations for fields within the footprint of the main GALAH survey that contain at least 225 stars from {\it Tycho-2} in the range $9 < V_{\rm JK} < 12$. These configurations are suggested by the \textsc{ObsManager} software for observation during evening and morning twilight. Because these stars are brighter than GALAH survey targets, the standard exposure times are shortened to $3 \times 6$ minutes instead of $3 \times 20$ minutes. As of 30 January 2016 we have observed 4448 {\it Tycho-2} stars in 26 of these targeted fields. An additional 6586 stars from the {\it Tycho-2} catalogue have been observed as part of our regular survey fields. 

Although we do not know exactly which {\it Tycho-2} stars will be included in {\it Gaia} DR1 or TGAS, we have made a portion of the current GALAH derived quantities for {\it Tycho-2} stars publicly available ahead of the first {\it Gaia} data release, which will take place on 14 September 2016. Our goal in publishing this GALAH-TGAS catalogue is to facilitate the exploitation of {\it Gaia} DR1 and to demonstrate the quality of GALAH derived quantities with a data set that will be extremely well studied in the near future. Table \ref{tgas} lists ID numbers from GALAH, {\it Tycho-2} and 2MASS, right ascension and declination from UCAC4, T$_{\rm eff}$, log(g), [Fe/H], [$\alpha$/Fe], radial velocity, distance modulus and $E(B-V)$ reddening for the first ten stars in the catalogue; the full table is available on the Vizier catalogue service. The full table includes analysis results for 3801 observations of 3675 {\it Tycho-2} stars in targeted fields and 6879 observations of 6185 serendipitiously observed {\it Tycho-2} stars for which we have successfully determined stellar parameters.

\begin{table}
\tabcolsep=0.11cm
\caption{The GALAH-TGAS catalogue. The full catalogue is available online; a portion of the table is published here for guidance as to form and content.}
\label{tgas}
\scalebox{0.9}{
\begin{tabular}{cccccccccccc}
  \hline
GALAH ID & Tycho-2 ID & 2MASS ID & $\alpha$ & $\delta$ & T$_{\rm eff}$ (K) & log(g) & [Fe/H] & [$\alpha$/Fe] & $v_{\rm rad}$ (km s$^{-1}$) & $(m-M)_V$ & $E(B-V)$ \\
  \hline
22245 & 9512-01937-1 & J13593210-8450329 & 13h59m32.11s & -84d50m32.9s & 6411 & 3.98 & -0.35 & 0.03 & 19.422 & 9.075 & 0.063 \\
26942 & 9508-02667-1 & J13373358-8420456 & 13h37m33.58s & -84d20m45.6s & 5832 & 4.07 & -0.32 & -0.02 & 39.962 & 8.094 & 0.022 \\
27265 & 9509-01044-1 & J14203533-8418486 & 14h20m35.34s & -84d18m48.6s & 6162 & 3.94 & -0.55 & 0.01 & 34.000 & 8.141 & 0.083 \\
28459 & 9508-02321-1 & J13383428-8411192 & 13h38m34.28s & -84d11m19.2s & 6006 & 4.21 & -0.28 & 0.02 & 8.263 & 7.934 & -0.022 \\
28516 & 9508-02638-1 & J13534492-8410566 & 13h53m44.92s & -84d10m56.7s & 4207 & 1.73 & -0.56 & 0.39 & 32.238 & 11.894 & 0.088 \\
29248 & 9509-00704-1 & J14072550-8406074 & 14h07m25.51s & -84d06m07.4s & 5813 & 4.45 & -0.00 & -0.04 & 22.343 & 7.596 & 0.087 \\
35133 & 9509-02342-1 & J14031173-8332020 & 14h03m11.73s & -83d32m02.1s & 5086 & 3.53 & -0.38 & 0.13 & -23.160 & 9.190 & 0.052 \\
35890 & 9508-02273-1 & J13594214-8327377 & 13h59m42.14s & -83d27m37.8s & 4858 & 2.72 & -0.01 & 0.13 & -16.954 & 11.006 & 0.046 \\
36337 & 9508-01621-1 & J13463498-8325184 & 13h46m34.99s & -83d25m18.5s & 5763 & 4.21 & -0.46 & 0.12 & 1.803 & 7.134 & 0.056 \\
50312 & 9440-00171-1 & J15071754-8214018 & 15h07m17.54s & -82d14m01.9s & 6000 & 3.72 & -0.20 & 0.03 & 5.900 & 8.363 & 0.170 \\
  \hline
\end{tabular}}
\end{table}

Barycentric-corrected radial velocities are determined through cross-correlation against a grid of AMBRE model spectra \citep{dLRB12}, as described in \citet{K16}. We use the HERMES blue, green, and red arm spectra for radial velocity determination, but not the IR arm spectra due to a relative lack of stellar features and a large number of telluric features. Adopted radial velocities are the mean of the values in the three arms, and the reported uncertainty is the standard deviation. Note that if the radial velocity measured from one arm is notably discrepant, e.g., is further from the mean than two times the difference between the measurements from the other two arms, it is excluded from the final radial velocity estimate. 98\% of all of our GALAH stars have a standard deviation of less than 0.6~km s$^{-1}$. 

The typical error on the radial velocity combined from the measurements in the three arms is a combination of systematic errors. One main contributor is the uncertainty that comes from the wavelength calibration itself. Spectra have been wavelength calibrated using a spectrum of a Thorium-Xenon arc lamp. Xenon lines dominate these spectra and we had to calibrate their wavelengths from the HERMES spectra themselves because of a lack of reliable linelist information in the literature. The wavelength calibration is therefore only accurate to $0.1$ to $0.5$~km s$^{-1}$, as can be seen in Figure 12 of \citet{K16}. The systematic offset in radial velocity between different arms is very low on average, typically $-$0.16~km s$^{-1}$ between the green and blue arms and $-$0.25~km s$^{-1}$ between the red and blue arms. For any given star there is a $1 \sigma$ probability that the difference in radial velocity between any two arms will be as large as $0.5$ to $0.75$~km s$^{-1}$, depending on the arms. This can be seen in Figure 18 of \citet{K16}. 

\begin{figure} 
\resizebox{0.45\textwidth}{!}{\includegraphics{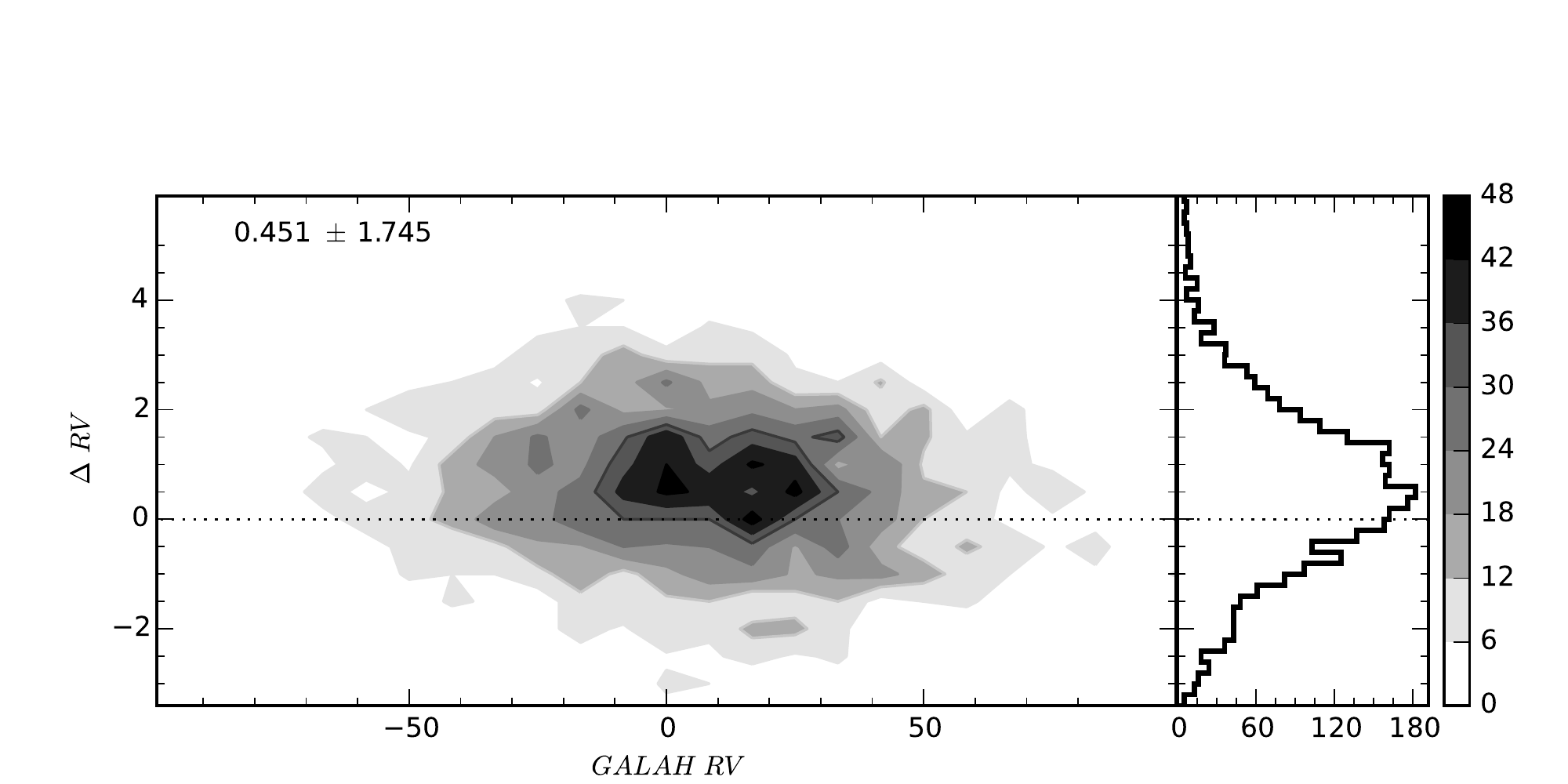} }\hspace{10pt} \resizebox{0.45\textwidth}{!}{\includegraphics{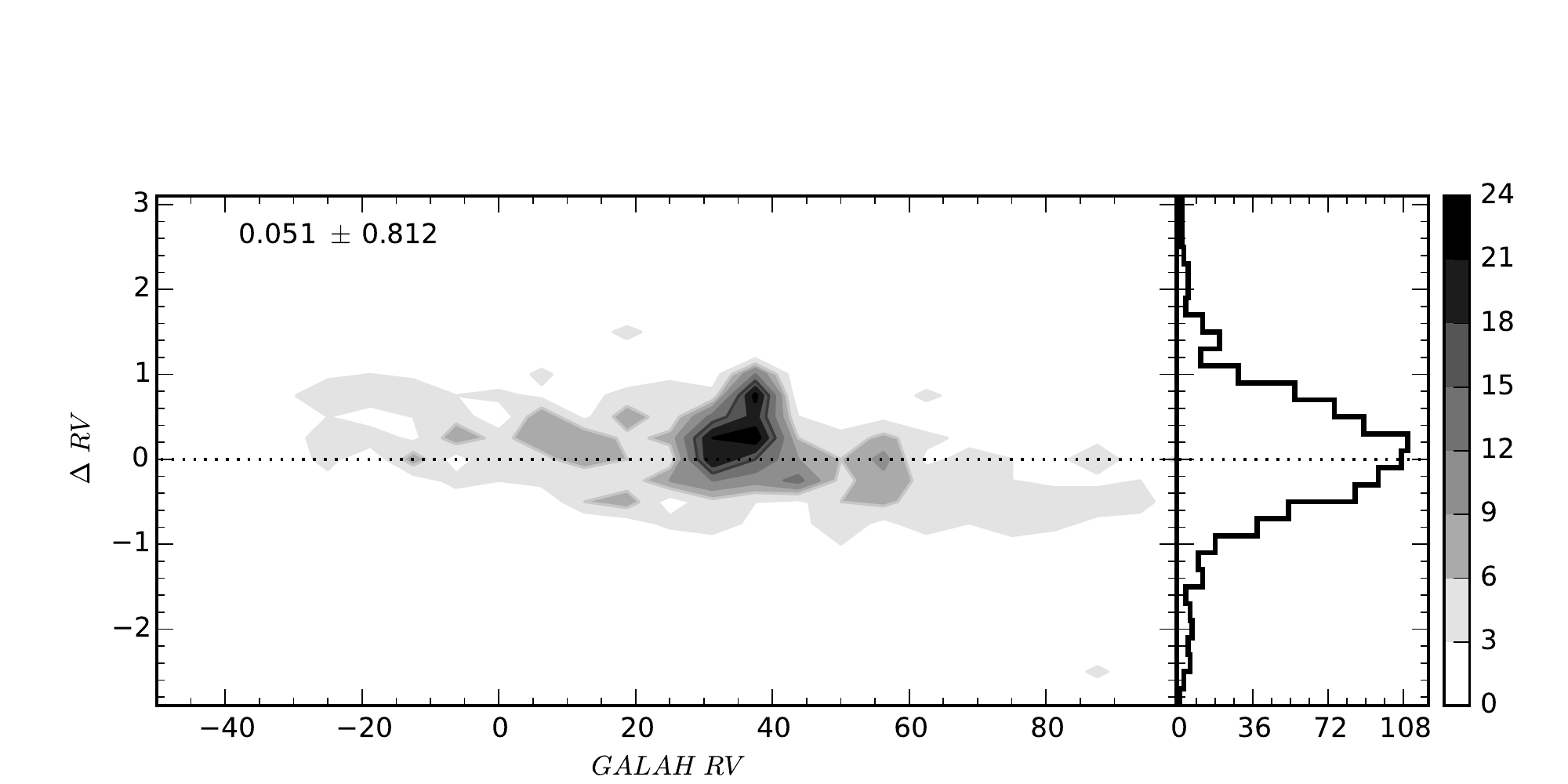} } 
\caption{A comparison of GALAH radial velocities with values from RAVE ({\it left panel}) and APOGEE ({\it right panel}). The x-axis is the GALAH radial velocity in km s$^{-1}$ while the y-axis is the difference between the GALAH value and that from the other survey, in the sense (GALAH-other). Color, as denoted by the bar, indicates the number density of stars. The right hand portion of each figure shows the distribution of the difference in radial velocity. The top left corner lists the mean and standard deviation in radial velocity difference. 
\label{fig:rv_survey_comp}}
\end{figure}

We have compared GALAH radial velocities to a number of sources in the literature. As presented in \citet{K16}, the values show good agreement with literature values for four clusters, M67, NGC~1851, NGC~288, and 47~Tuc. We can also verify our radial velocity accuracy by comparing GALAH values with those from other surveys. Through the end of January 2016 (the time period discussed in this study), there are 9388 and 664 targets that have also been observed by RAVE and APOGEE (RAVE DR4: \citealt{KGS13}; SDSS DR10: \citealt{Ahn14}), respectively. As the survey continues, this will prove to be an invaluable sample for database cross-comparison. Currently, it provides a useful comparison set for our radial velocities. The left panel of Figure\,\ref{fig:rv_survey_comp} shows the distribution of the difference in velocities between RAVE and GALAH. For this comparison, we have trimmed the GALAH-RAVE overlap sample to 3434 stars, based on the following quality criteria from the RAVE catalogue (M. Steinmetz, priv. comm.): 
\begin{itemize} 
\item logg$\_$K $>$ 0.5 dex
\item SNR$\_$K $>$ 20
\item eHRV $<$ 10 km/s
\item Teff$\_$K $>$ 4000 K
\item CHISQ$\_$c $<$ 2000 
\item c1, c2, c3 $=$ n
\item Algo$\_$Conv$\_$K $=$ 0 
\end{itemize}
 
The mean offset between GALAH and RAVE is 0.45 km s$^{-1}$ with a standard deviation of 1.75 km s$^{-1}$. Since RAVE uses lower resolution spectra than GALAH and reports a typical radial velocity uncertainty of 2 km s$^{-1}$, this is very good agreement. APOGEE is also consistent with GALAH, showing a mean offset of 0.05$\pm$0.81 km s$^{-1}$ (Figure\,\ref{fig:rv_survey_comp}, right panel). 

For the stellar parameter determination, we use a combination of the spectral synthesis program Spectroscopy Made Easy (SME) (\citealt{VP96}; \citealt{PV16}) and the data-driven \textit{Cannon} by \citet{NHR15a}. This approach delivers both accurate and precise parameters and is computationally inexpensive, as \textit{The Cannon} takes only 0.13 seconds to compute seven stellar labels for one spectrum.

We first use SME to determine stellar parameters and [$\alpha$/Fe] abundances for a subset of 2576 GALAH stars spanning the entire range of parameters covered by the survey. This subsample is then used as the training data set for \textit{The Cannon}. To obtain the highest precision and accuracy, this representative training set is comprised of only high quality spectra (SNR $>$ 95 per resolution element), as well as high-fidelity validation targets including observed benchmark stars with reliable, independent stellar parameters \citep{HJG15}, well studied open and globular cluster stars, and stars with confirmed asteroseismic surface gravities.

\begin{figure}
\resizebox{0.45\textwidth}{!}{\includegraphics[page=1]{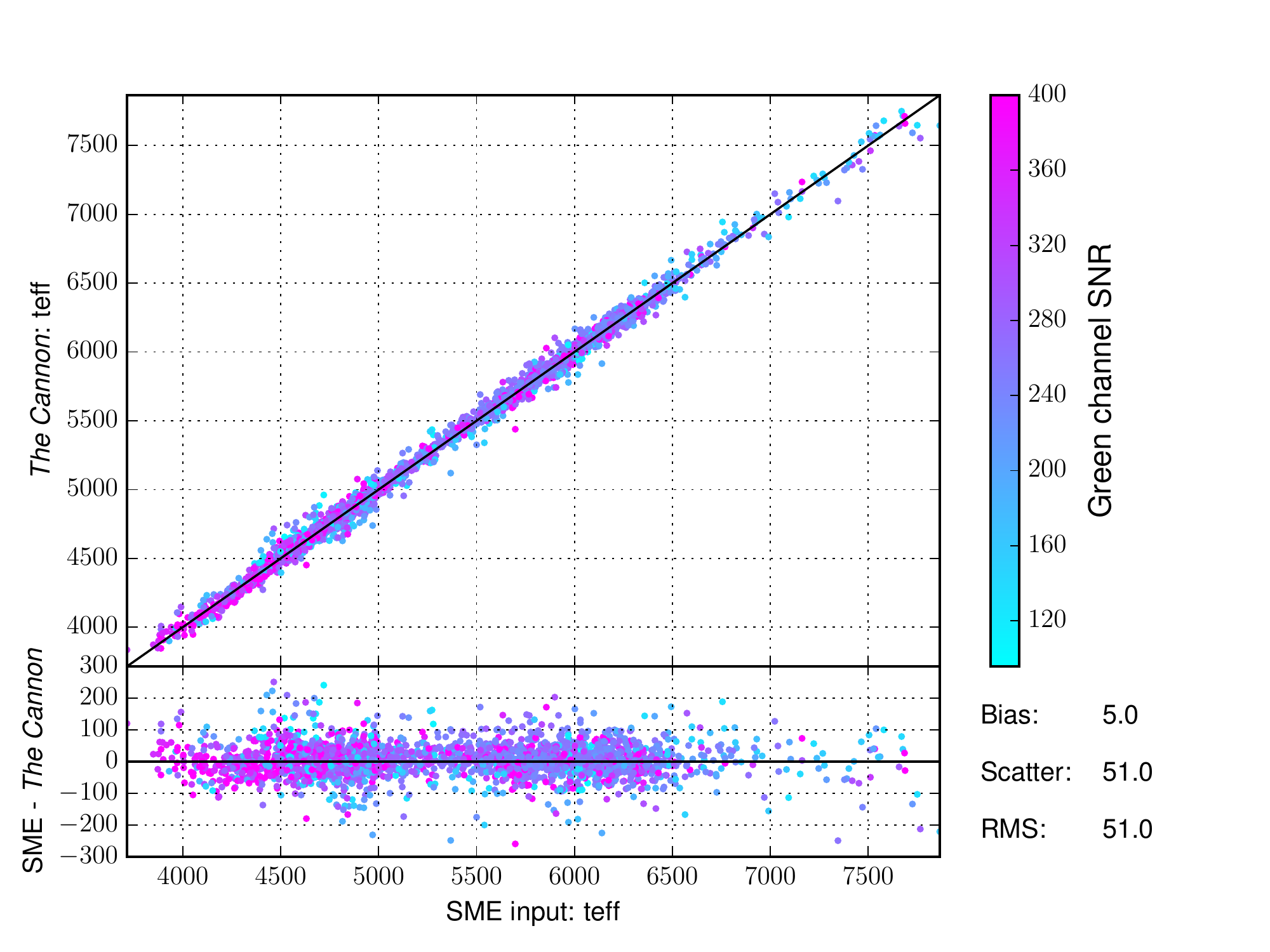} }
\resizebox{0.45\textwidth}{!}{\includegraphics[page=2]{Cannon_iDR1_trainingset_v_a_leave_out_test_cut400.pdf} }
\resizebox{0.45\textwidth}{!}{\includegraphics[page=3]{Cannon_iDR1_trainingset_v_a_leave_out_test_cut400.pdf} }
\caption{Results from 20\% leave-out cross validation tests for T$_{\rm eff}$ (upper left), log(g) (upper right) and [Fe/H] (lower left), as described in the text. These tests were repeated five times, and data from all five tests are plotted together in this figure, colour-coded by the green channel signal to noise ratio per resolution element. For each parameter, the upper panel shows {\textit Cannon} versus SME parameters with a 1:1 correspondence line drawn in solid black, and the lower panel shows the difference between the two as a function of the SME values.
\label{ctest}}
\end{figure}

Our initial estimates for T$_{\rm eff}$, log(g) and [Fe/H] are determined through cross-correlation against a grid of AMBRE model spectra that is larger and more finely sampled than the grid used for radial velocity determination, as decribed in \citep{K16}. SME takes these as input and determines the stellar parameters by fitting synthetic spectra to observations, returning optimal parameters corresponding to the minimum $\chi^2$ \citep{PV16}. The SME synthesis employs MARCS model atmospheres \citep{GEE08}, and includes NLTE corrections for Fe \citep{L12}. The global parameters T$_{\rm eff}$, log(g), [Fe/H], $v_{\rm mic}$, $V_{\sin i}$, and $v_{\rm rad}$ are optimized for unblended lines in the spectra, including the $\mathrm{H}\alpha$, $\mathrm{H}\beta$, FeI/II, ScI/II, and TiI/II lines which have reliable atomic data. The optimal global parameters returned by SME are subsequently fixed, and an error weighted [$\alpha$/Fe] is calculated from $\chi^2$ optimization for selected lines of $\alpha$-process elements Mg, Si, and Ti.

\textit{The Cannon} then uses the normalised spectra, SME-determined stellar parameters and $\alpha$-abundances as \textit{labels} for the reference set of stars and generates a spectral model of the GALAH spectra at rest-frame wavelength. This generative \textit{Cannon} model relates the observed flux to the labels provided (the training step) and is used to determine those same labels for all stars in the survey. We find that a second order polynomial model works well for GALAH spectra. In addition to the SME stellar parameters and abundances, we also include extinction values as a label for \textit{The Cannon}, allowing it to take into account the effect of diffuse interstellar bands on some $\alpha$-element lines, and thus providing a more accurate final $\alpha$-abundance. The extinction, $A_K$, is derived as described by \citet{ZJ13}, using the 2MASS $H$-band and WISE 4.5 $\mu$m photometry (\citealt{SCS06}; \citealt{WEM10}). For each star, \textit{The Cannon} delivers a set of seven labels consisting of: T$_{\rm eff}$, log(g), [Fe/H], $v_{\rm mic}$, $V_{\sin i}$, [$\alpha$/Fe] and $A_K$.  Figure \ref{ctest} shows the results of 20\% leave-out cross-validation tests demonstrating that \textit{The Cannon} is well able to determine the stellar labels to high precision. This test involves omitting a random 20\% of the training set, then comparing the parameter values predicted for those omitted stars by \textit{The Cannon} with the values determined using SME. We find the following biases and precisions: $\Delta$ T$_{\rm eff} = 5 \pm 41\,\mathrm{K}$, $\Delta log(g) = 0.01 \pm 0.17\,\mathrm{dex}$, $\Delta \mathrm{[Fe/H]} = 0.005 \pm 0.056\,\mathrm{dex}$. We have provided here a summary of the GAALH spectroscopic analysis pipeline, the details of which will be given in Asplund et al. (in prep).

Distances are determined using theoretical isochrones, as discussed in \citet{ZMB10}, assuming that each star undergoes a standard stellar evolution and that its spectrum shows no peculiarities. The latter is checked by  morphological classification of spectra which is based on a t-distributed stochastic neighbor embedding algorithm (Traven et al. 2016, in preparation;  for description of the algorithm see \citealt{vdM13} and references therein). Absolute magnitudes in Johnson $V$ and 2MASS $J$ bands are estimated from theoretical Padova isochrones \citep{BGM08} with weights determined (as described in \citealt{ZMB10}) using a mass function from \citet{C03}, a flat prior on ages between 0.5 and 12 Gyr and a flat prior on space density. Stellar parameter values determined from GALAH spectra by the Cannon algorithm \citep{NHR15a} are assumed to have a standard deviation of 100~K in temperature, 0.25 dex in gravity and 0.1 dex in metallicity. These error estimates are compatible with differences between parameter values determined by GALAH and APOGEE \citep{HSJ15} for stars observed by both surveys. 

\begin{figure*}
\resizebox{0.45\textwidth}{!}{\includegraphics{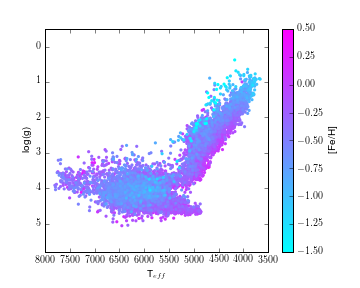}}\hspace{10pt} 
\resizebox{0.45\textwidth}{!}{\includegraphics{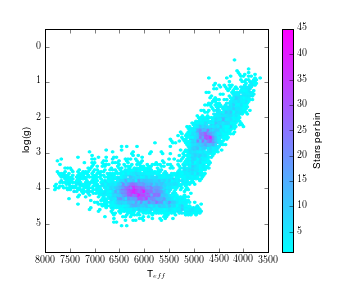}}\hspace{10pt} 
\caption{Stellar parameters T$_{\rm eff}$ and log(g) for stars in the GALAH-TGAS catalogue, colour-coded by metallicity (left panel), and binned into hexagons and colour-coded by the number of stars per bin (right panel).}
 \label{tgf}
\end{figure*}

Comparison of absolute magnitudes with the apparent $V$ magnitude from the latest version of the APASS survey \citep{HM13} and $J$ magnitude from 2MASS \citep{CSv03} leads to an estimate of the distance modulus as well as reddening along the line of sight, if standard relations $A_V = 3.1 E(B-V)$ and $A_J = 0.887 E(B-V)$ are used. The typical accuracy of derived distance modulus is 0.4 mag (implying a distance uncertainty of $\sim 20$\%) and the typical accuracy of colour excess is $\sim 0.04$~mags. Such errors apply to main sequence (MS) and to red giant branch  stars, but for the transition region between the MS turn-off and the red giant branch the errors increase considerably. Comparison of our and literature values of distance moduli and reddenings for members of three open clusters (NGC~2243, Pleiades and NGC~2516) and one globular cluster (NGC~6362) confirm such error estimates. GALAH targets are located at least ten degrees from the Galactic plane, so uncertainties in reddening do not affect derived values of distance modulus significantly. This is confirmed by a median value of just 0.03~mag for the colour excess. Here we publish spectrophotometric distances for {\it Tycho-2} stars, which are much brighter and so closer than typical stars observed by GALAH, where MS stars are generally within 1~kpc from the Sun and red clump stars are at distances of $\sim 3$~kpc.  

\begin{figure*}
\resizebox{0.45\textwidth}{!}{\includegraphics{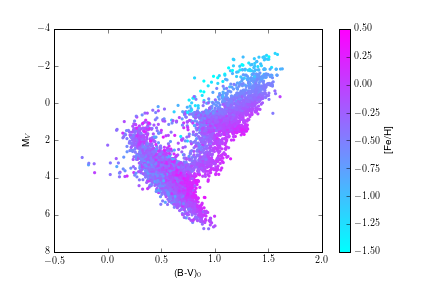}}\hspace{10pt}
\resizebox{0.45\textwidth}{!}{\includegraphics{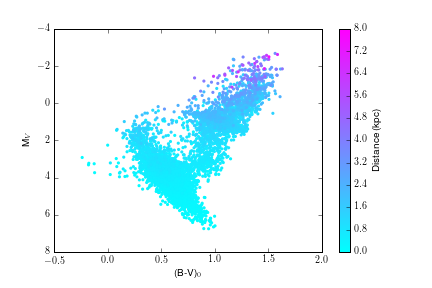}}\hspace{10pt}
 \caption{Absolute colour-magnitude diagram for stars in the GALAH-TGAS catalogue, colour-coded by metallicity (left panel) and distance (right panel).}
 \label{abs_cmd}
\end{figure*}

The derived stellar parameters for {\it Tycho-2} stars observed by GALAH appear to be quite reasonable. This can be seen in Figure \ref{tgf}, which shows effective temperature versus surface gravity, colour-coded by metallicity (left panel) and binned into hexagons and colour-coded by the number of stars per bin (right panel). As one might expect, there is a clear gradient in metallicity across the red giant branch and the upper main sequence. There are few metal-poor stars ([Fe/H]$< -1.5$), and the majority of the stars are on the main sequence, indicating that these stars belong almost entirely to the Galactic disk.

Figure \ref{abs_cmd} shows colour-magnitude diagrams in dereddened $(B-V)$ and absolute $M_{\rm V}$, colour-coded by metallicity (upper left panel) and distance (upper right panel), and binned into hexagons and colour-coded by the number of stars per bin (lower left panel). 

\begin{figure*}
\resizebox{0.45\textwidth}{!}{\includegraphics{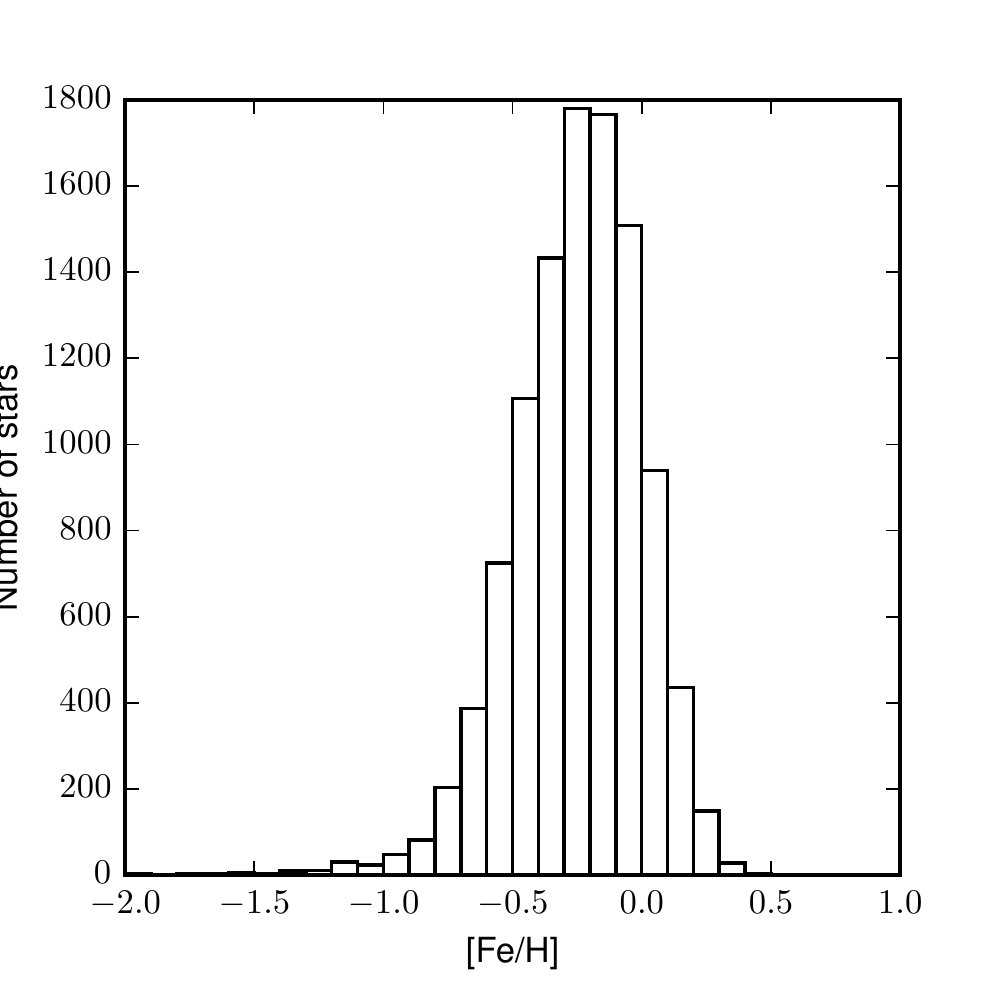}}\hspace{10pt}
\resizebox{0.45\textwidth}{!}{\includegraphics{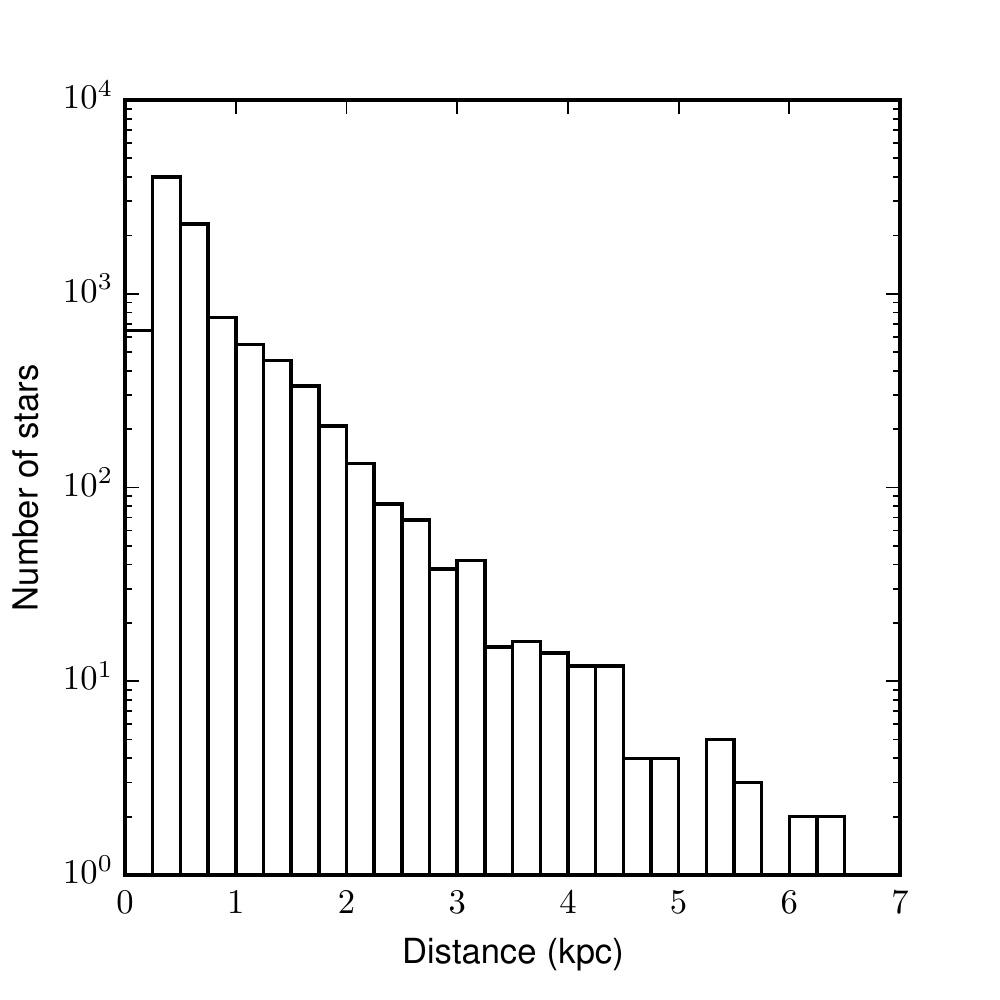}}\hspace{10pt}
 \caption{Histograms of metallicity (left panel) and distance (right panel). The distance histogram is shown on a logarithmic scale to enhance the visibility of stars at larger distance, since the majority of stars in the GALAH-TGAS catalogue are quite nearby.}
 \label{feh_dist}
\end{figure*}

\begin{figure*}
\resizebox{0.8\textwidth}{!}{\includegraphics{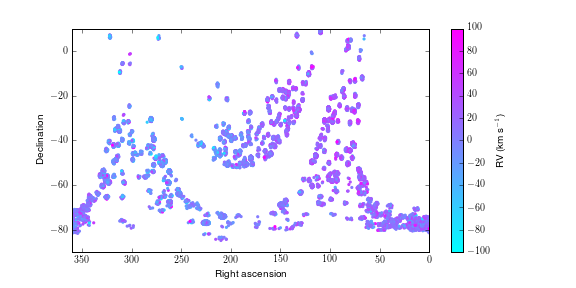}}
 \caption{Map of the GALAH-TGAS catalogue in right ascension and declination, colour-coded by radial velocity. The Solar motion relative to the Local Standard of Rest can be clearly seen.}
 \label{lbrv}
\end{figure*}

Since these stars are fairly bright, their distribution across the Milky Way is somewhat limited relative to the full GALAH survey. This can be seen in Figure \ref{feh_dist}, which shows histograms of metallicity (left panel) and distance (right panel). Although the stars in the GALAH-TGAS catalogue do span the full sky coverage of the GALAH survey (as can be seen in Figure \ref{lbrv}), they are mainly members of the thin disk: they have relatively high metallicities and are located within 2 kpc of the Sun. 

\section{Summary}
The GALAH Survey has made significant progress toward its goal of observing one million stars in the Milky Way over its first two years of survey observing. Up to 30 January 2016 we have observed 209,345 stars in the main survey, 845 targeted stars in globular and open clusters, 2,218 stars in the CoRoT anticentre fields, and 9,847 stars for the thin-thick disk program during the Pilot Survey, and an additional 31,365 stars have been observed by the K2-HERMES program. 

We have also intentionally observed 4448 Tycho-2 stars in 26 fields to correspond with the first {\it Gaia} data release, with another 6586 stars observed serendipitously in the regular GALAH Survey fields. Of these, we are making available analysis results for 10680 observations of 9860 stars (3801 observations of 3675 targeted stars and 6879 observations of 6185 serendipitous stars) that have successfully been processed through our parameter and abundance determination pipeline. A catalogue of stellar parameters, radial velocities, distance moduli and reddening for these successfully analysed stars is presented in this publication, to support broad scientific exploitation of the first {\it Gaia} data release. As demonstrated above, these parameters look quite robust. We anticipate that they will improve further when we adapt our spectroscopic analysis pipeline to include the stellar distances derived by the Tycho-Gaia Astrometric Solution \citep{MLH15} and future {\it Gaia} data releases. Combining spectroscopic datasets with {\it Gaia} data serves many important purposes beyond improving spectroscopic analysis. Future GALAH data releases will add elemental abundance information for the stars with the best {\it Gaia} parallaxes and proper motions, enabling chemodynamic studies in the Solar neighbourhood and throughout the Galaxy, and adding kinematic information into chemical tagging.

The target selection and field tiling for GALAH are fixed, and we will continue to follow the same observing rules for the duration of the survey, maintaining our straightforward selection function. Based on Galactic models and our target selection strategy we anticipate a final data set that is dominated by the thin and thick disks, but despite the small fraction of halo and bulge stars expected ($<1\%$), these data sets will also have significant scientific value. Further details on the data set as observed will be available in Sharma et al. (in prep). 


\section*{Acknowledgments}

SLM and DBZ acknowledge support from Australian Research Council grants DE140100598 and FT110100743. JPM is supported by a UNSW Vice-Chancellor's Research Fellowship. K.L. and S.B. acknowledge funds from the Alexander von Humboldt Foundation in the framework of the Sofja Kovalevskaja Award endowed by the Federal Ministry of Education and Research as well as funds from the Swedish Research Council (Grant nr. 2015-00415\_3) and Marie Sklodowska Curie Actions (Cofund Project INCA 600398).This work was partly supported by the European Union FP7 programme through ERG grant number 320360.

\bibliography{obs-overview}

\end{document}